\documentclass[letterpaper,paper,11pt]{AAS}

\usepackage{bm}
\usepackage{amsmath}
\usepackage{amssymb}
\usepackage{amsfonts}
\usepackage{graphicx}
\graphicspath{{figures/}}
\usepackage[colorlinks=true,pdfstartview=FitV,
            linkcolor=black,citecolor=black,urlcolor=black]{hyperref}
\usepackage{overcite}
\usepackage{footnpag}
\usepackage{siunitx}
\usepackage{booktabs}
\usepackage{caption}
\usepackage{subcaption}
\DeclareCaptionFont{aasfont}{\fontsize{10}{10}\selectfont\bfseries\boldmath}
\usepackage{threeparttable}
\usepackage{multirow}
\usepackage{balance}
\usepackage{placeins}
\usepackage{float}

\newcommand{\Bbold}{\boldsymbol{B}}

\newcommand{\Ibold}{\boldsymbol{I}}

\newcommand{\xbold}{\boldsymbol{x}}

\let\AASoriginalthebibliography\thebibliography
\renewcommand{\thebibliography}[1]{%
  \AASoriginalthebibliography{#1}%
  \fontsize{9}{9}\selectfont
}

\PaperNumber{XX-XXX}

\begin{document}

\title{High-Fidelity Baseline Design and Station Keeping Analyses for
Earth--Moon Vertical Orbits}

\author{Derek Phung\thanks{Undergraduate Student, Mechanical and Aerospace
Engineering, University of California, Irvine, California, 92617.}
\ and Yuri Shimane\thanks{Assistant Professor, Mechanical and Aerospace
Engineering, University of California, Irvine, California, 92617.}}

\maketitle

\begin{abstract}
Earth--Moon vertical orbits are periodic solutions of the circular restricted three-body problem (CR3BP) with substantial out-of-plane motion, considered for high-latitude lunar observation and cislunar navigation. This work develops an end-to-end framework for high-fidelity reference generation under rotating-frame geometry constraints and receding-horizon model predictive control. A representative $L_2$ vertical orbit yields a 5.5-year reference with negligible deterministic correction and remains maintainable in five-year Monte Carlo simulations under realistic uncertainties. The workflow produces locally optimal solutions across 90 consecutive daily start epochs and 200 catalog cases. Finite-Time Lyapunov Exponent along the reference and stable/unstable CR3BP directions interpret station-keeping behavior.
\end{abstract}

\section{Introduction}

Periodic libration-point orbits (LPOs) provide several distinct in-plane and
out-of-plane geometries in the Earth--Moon system and are widely used as
initial designs for cislunar trajectories
~\cite{GomezMondelo2001,HouLiu2009,JPLPeriodicOrbits}.
Vertical orbits are of particular interest because their out-of-plane motion can
support high-latitude lunar observation and has also been considered in
cislunar navigation architectures. Grebow et al. examined multibody
architectures using halo, vertical, and other $L_1/L_2$ families for lunar
South Pole coverage~\cite{GrebowEtAl2008}. Pergola and Alessi characterized
$L_1$ vertical Lyapunov, halo, and Lissajous orbits using lunar surface
coverage, eclipse, and orbit-maintenance metrics~\cite{PergolaAlessi2012},
while Heiligers et al. demonstrated continuous observation of the lunar South
Pole and Aitken Basin using a two-spacecraft constellation of solar-sail $L_2$
vertical Lyapunov orbits~\cite{HeiligersEtAl2018}. Zhang and Xu also considered
vertical Lyapunov orbits within Earth--Moon libration-point navigation
constellations and reported improved cislunar navigation performance for
architectures with the $L_1$ navigation satellite on a halo or vertical
Lyapunov orbit~\cite{ZhangXu2015}. 

A circular restricted three-body problem (CR3BP) design alone does not establish how a given LPO will behave after
transition to a high-fidelity ephemeris model. Directly mapping a CR3BP initial
condition to a selected physical epoch and propagating it in that model does
not by itself ensure a long-duration trajectory that retains the intended
CR3BP geometry. Variation in the Earth--Moon distance, solar gravity, the
selected initial epoch, and the transition procedure can all change the
resulting trajectory~\cite{ParkHowell2024Transition}. A dynamically consistent
reference is also important for station keeping: if the nominal trajectory is
inconsistent with the propagation model, the controller can spend control
effort correcting reference-model error rather than operational
dispersions~\cite{HowellPernicka1993}. High-fidelity baseline optimization
under rotating-frame geometry constraints provides one approach for
constructing such dynamically consistent
references~\cite{ShimaneHo2025Baseline}.

Several station-keeping approaches have been developed for LPOs
\cite{ShirobokovEtAl2017}. Target-point and $x$-axis-crossing methods compute a
correction that drives one or more downstream conditions back toward the
reference~\cite{HowellPernicka1993}. Floquet-mode methods instead use the
linear structure of the periodic orbit and suppress components associated with
the unstable mode~\cite{GomezEtAl1998}. More recent optimization-based
extensions include phase-constrained $x$-axis crossing control
\cite{ShimaneHoWeiss2025PhaseXAC} and full-state model predictive control (MPC)
that repeatedly solves a finite-horizon problem and executes only the first
planned maneuver~\cite{ShimaneEtAl2025OutputFeedback}. Once a station-keeping
method is selected, Monte Carlo analyses can be used to evaluate its robustness
under navigation uncertainty, maneuver-execution error, and spacecraft
disturbances~\cite{DavisEtAl2017,DavisEtAl2022,FoltaEtAl2014}.

Dynamical information can also provide insight into why a station-keeping
controller selects particular maneuver directions. Low-cost corrections have been linked
to local stable or restoring directions in operational and long-duration
studies~\cite{FoltaEtAl2014,PavlakHowell2012}, and finite-time stretching has
been used to interpret maneuver locations and targeting horizons near
Earth--Moon halo orbits~\cite{MuralidharanHowell2022}. In this work, we compute the finite-time Lyapunov exponent directly along the high-fidelity reference and, in a separate analysis, compare the planned MPC maneuvers with stable and unstable velocity directions evaluated along the
underlying periodic CR3BP orbit at matched orbit phases.

This work makes three main contributions. First, we extend high-fidelity baseline
optimization under rotating-frame geometry constraints to multi-year
Earth--Moon vertical orbit references, validate the resulting Spacecraft and
Planet Kernel (SPK) products, evaluate geometry-constraint sensitivity, and
test start-epoch robustness over 90 consecutive daily epochs. The same baseline-generation pipeline is also applied to 200 $L_1/L_2$
vertical orbits from the Jet Propulsion Laboratory (JPL) catalog to generate approximately six-month high-fidelity references. Second, we demonstrate long-duration receding-horizon station keeping along the
generated vertical orbits under navigation, maneuver-execution,
solar-radiation-pressure, and attitude-disturbance uncertainty, including a
five-year Monte Carlo analysis and a comparison across selected long-duration
$L_1$ and $L_2$ vertical orbits. Third, we interpret the MPC maneuver directions by combining the finite-time
Lyapunov exponent along the actual high-fidelity reference with a comparison between planned first MPC maneuvers and stable and unstable CR3BP velocity directions evaluated at matched orbit phases.

\begin{figure}[!htbp]
    \centering
    \includegraphics[width=0.90\textwidth]{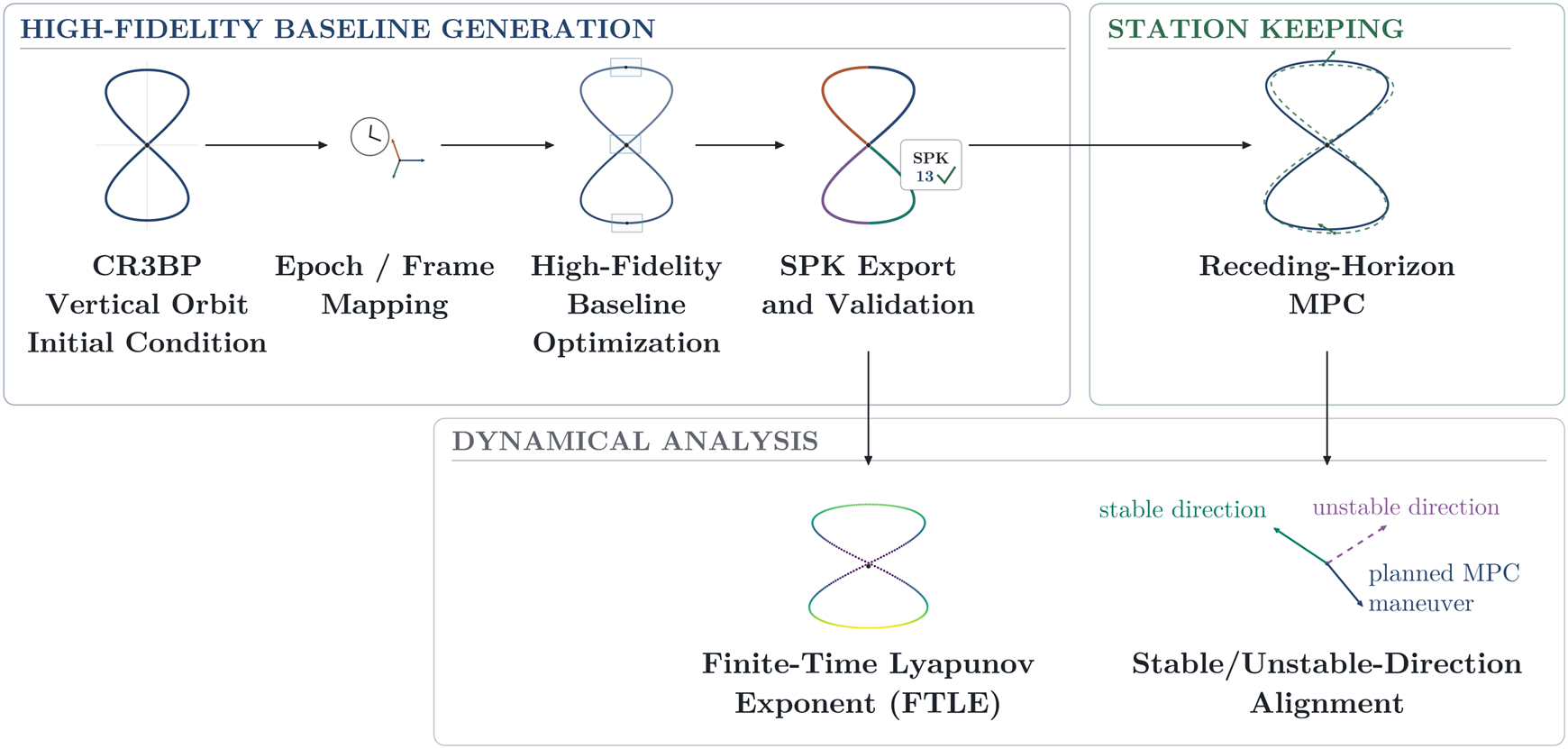}
    \caption{Workflow for high-fidelity baseline generation, station keeping, and dynamical analysis}
    \label{fig:workflow}
\end{figure}

Relative to prior studies that considered vertical orbits for
specific mission applications or selected trajectories, this work generates
high-fidelity references across a broader portion of the vertical orbit family,
including 200 $L_1/L_2$ catalog members. Selected multi-year references are then carried into closed-loop
receding-horizon station-keeping analyses under realistic uncertainties. Figure~\ref{fig:workflow} summarizes the end-to-end workflow.

The remainder of the paper is organized as follows. The High-Fidelity Baseline
Generation section presents the CR3BP initial conditions and epoch mapping, high-fidelity
dynamics, high-fidelity baseline optimization, and SPK export and validation. The
Station-Keeping and Dynamical Analysis section formulates the receding-horizon
controller, uncertainty and disturbance models, finite-time Lyapunov exponent,
and CR3BP stability-direction comparison. The Numerical Results section then presents detailed baseline-generation,
robustness, station-keeping, and dynamical analyses for a selected $L_2$
vertical orbit, followed by family-wide baseline-generation results and a
station-keeping comparison across selected long-duration $L_1$ and $L_2$
vertical orbits. The final section summarizes the principal conclusions.

\FloatBarrier

\section{High-Fidelity Baseline Generation}

This section describes how CR3BP catalog initial conditions are mapped to an initial epoch,
optimized in the high-fidelity ephemeris model under rotating-frame geometry constraints,
and exported as validated SPK references.

\subsection{CR3BP Initial Conditions and Epoch Mapping}

The vertical-orbit initial conditions are taken from the JPL Three-Body
Periodic Orbit Catalog\footnote{Access database at:
\url{https://ssd.jpl.nasa.gov/tools/periodic_orbits.html}}, which provides
CR3BP states, periods, family labels, and linear stability
information~\cite{JPLPeriodicOrbits}. The representative case used for the
detailed analyses is the $L_2$ vertical orbit with JPL catalog ID 1036,
referred to hereafter as $L_2$ 1036.

The Earth--Moon CR3BP is used to define these catalog initial
conditions~\cite{Szebehely1967,GomezMondelo2001,HouLiu2009}. The
spacecraft is modeled as an infinitesimal third body moving under the gravity
of the Earth and Moon, whose masses are $m_1$ and $m_2$, respectively. We use the standard Earth--Moon CR3BP normalization, with mass parameter
$\mu=m_2/(m_1+m_2)$. The production mapping uses the distance unit
$\mathrm{DU}=389{,}703$~km, the time unit $\mathrm{TU}=382{,}981$~s, and
$\mu=0.01215058560962404$. This normalization places the Earth and Moon at
$\begin{bmatrix}-\mu & 0 & 0\end{bmatrix}^{T}$ and
$\begin{bmatrix}1-\mu & 0 & 0\end{bmatrix}^{T}$, respectively, in the
barycentric Earth--Moon rotating frame. The normalized frame angular velocity is
$\bm{\omega}=[0\ 0\ 1]^T$.

Let $\bm{r}=[x\ y\ z]^T$ denote the spacecraft position with respect to the
Earth--Moon barycenter and let
$\bm{v}=\dot{\bm{r}}=[v_x\ v_y\ v_z]^T$ denote its velocity relative to the
rotating frame. The state is
$\bm{x}=[\bm{r}^{T}\ \bm{v}^{T}]^{T}$. Define the Earth-to-spacecraft and
Moon-to-spacecraft position vectors
\begin{equation}
    \bm{r}_1
    =
    \begin{bmatrix}
        x+\mu &
        y & 
        z
    \end{bmatrix}^T,
    \qquad
    \bm{r}_2
    =
    \begin{bmatrix}
        x-1+\mu &
        y &
        z
    \end{bmatrix}^T,
    \label{eq:cr3bp_primary_vectors}
\end{equation}
with $r_1=\|\bm{r}_1\|_2$ and $r_2=\|\bm{r}_2\|_2$. In canonical units, the
CR3BP dynamics are
\begin{equation}
    \dot{\bm{x}}
    =
    \begin{bmatrix}
        \bm{v} \\
        -\dfrac{1-\mu}{r_1^3}\bm{r}_1
        -\dfrac{\mu}{r_2^3}\bm{r}_2
        -\bm{\omega}\times(\bm{\omega}\times\bm{r})
        -2\bm{\omega}\times\bm{v}
    \end{bmatrix}.
    \label{eq:cr3bp}
\end{equation}
Each catalog initial condition lies on a periodic CR3BP orbit with
period $T_{\mathrm{C}}$, such that
$\bm{x}_{\mathrm{C}}(t+T_{\mathrm{C}})=\bm{x}_{\mathrm{C}}(t)$.
The periodic solution is repeated in phase to initialize the long-duration
nodal trajectory rather than being used directly as the final high-fidelity
reference.

At a selected physical epoch, the repeated CR3BP state is first translated from
the barycentric rotating origin to a Moon-centered Earth--Moon rotating origin
and dimensionalized using the CR3BP \(\mathrm{DU}\) and \(\mathrm{TU}\). We then
apply the full epoch-dependent $6\times6$ state transformation from the
Moon-centered Earth--Moon rotating frame to the Moon-centered J2000 inertial
frame. The rotating frame is defined using the Spacecraft, Planet, Instrument,
C-matrix, and Events (SPICE) system as a two-vector frame with its $x$-axis
aligned with the instantaneous Earth-to-Moon position vector and its
$y$-axis constructed from the instantaneous Earth-to-Moon relative velocity
vector, orthogonalized with respect to the $x$-axis. The $z$-axis completes the
right-handed orthonormal triad. The mapped states initialize the high-fidelity
multiple-shooting nodes. Rotating-frame geometry constraints and the trajectory
plots reported below use \(\mathrm{DU}=389{,}703\)~km.

\subsection{High-Fidelity Ephemeris Model}

The high-fidelity ephemeris model (HFEM) propagates the spacecraft in
Moon-centered J2000 using JPL ephemerides, physical parameters, and frame
definitions~\cite{Acton1996}. The natural dynamics are
\begin{equation}
    \dot{\bm{x}}
    =
    \bm{f}(\bm{x},t)
    =
    \begin{bmatrix}
        \bm{v} \\
        -\dfrac{\mu_M}{\|\bm{r}\|_2^3}\bm{r}
        +\displaystyle\sum_{b\in\mathcal{B}}\bm{a}_{b}
        +\bm{a}_{\mathrm{SH}}
        +\bm{a}_{\mathrm{SRP}}
    \end{bmatrix},
    \label{eq:hfem}
\end{equation}
where $\mu_M$ is the lunar gravitational parameter,
$\mathcal{B}$ is the set of perturbing bodies,
$\bm{a}_{b}$ is the third-body acceleration due to body $b$, and
$\bm{a}_{\mathrm{SH}}$ is lunar spherical-harmonic gravity evaluated in the
Moon's principal-axes frame; standard expressions for these accelerations are
used~\cite{Vallado2013}. The term $\bm{a}_{\mathrm{SRP}}$ is the perturbation
due to solar radiation pressure (SRP), assuming the standard cannonball
model~\cite{Vallado2013},
\begin{equation}
    \bm{a}_{\mathrm{SRP}}
    =
    P_0 C_R \frac{A}{m}
    \left(\frac{\mathrm{AU}}{d_{\odot}}\right)^2
    \hat{\bm{r}}_{\odot\rightarrow\mathrm{sc}},
    \label{eq:srp}
\end{equation}
where $P_0$ is the solar radiation pressure at 1 astronomical unit (AU),
$C_R$ is the radiation-pressure coefficient, $A/m$ is the spacecraft
area-to-mass ratio,
$d_{\odot}$ is the instantaneous Sun--spacecraft distance, and
$\hat{\bm{r}}_{\odot\rightarrow\mathrm{sc}}$ is the unit vector directed from
the Sun toward the spacecraft. For station keeping, the nominal cannonball parameters are
$C_R=1.15$, $A/m=0.016$~m$^2$/kg, and
$P_0=4.56\times10^{-6}$~N/m$^2$. Baseline generation uses the deterministic gravitational model with lunar
harmonics through degree and order 4 and no SRP. The station-keeping prediction
and truth models include continuous cannonball SRP without Earth or Moon
shadowing.

\subsection{High-Fidelity Baseline Optimization}

The high-fidelity baseline optimization follows the formulation of Shimane
and Ho~\cite{ShimaneHo2025Baseline}. Each CR3BP revolution is divided into four
coast arcs, such that for $N_{\mathrm{rev}}$ revolutions, $N_{\mathrm{node}}=4N_{\mathrm{rev}}+1$ with time interval between each node $\Delta t=\frac{T_{\mathrm{C}}}{4}$. Let $\bm{x}_k$ be the pre-impulse state at node $k$ and
$\Delta\bm{v}_k$ the impulsive correction. The nonlinear optimal control problem (OCP) is
\begin{subequations}
\begin{align}
    \min_{\bm{x}_k,\Delta\bm{v}_k,\sigma_k}\quad&
    \sum_{k=0}^{N_{\mathrm{node}}-1}\sigma_k
    \label{eq:baseline_objective}\\
    \mathrm{s.t.}\quad&
    \bm{x}_{k+1}
    =
    \bm{x}_k+\bm{B}\Delta\bm{v}_k
    +\int_{t_k}^{t_{k+1}}
    \bm{f}\!\left(
        \xbold(t),
        t\right)\,\mathrm{d}t,
    \label{eq:baseline_dynamics}\\
    &
    \|\Delta\bm{v}_k\|_2\leq\sigma_k\leq u_{\max},
    \label{eq:baseline_control}\\
    &
    \bm{c}_{s(k)}-\bm{\Delta}_{s(k)}
    \leq
    \bm{T}^{\mathrm{EM}}_{\mathrm{Inr}}(t_k)
    \bm{E}\bm{x}_k
    \leq
    \bm{c}_{s(k)}+\bm{\Delta}_{s(k)},
    \label{eq:baseline_geometry}
\end{align}
\end{subequations}
where $\Bbold = [\bm{0}_{3\times 3}; \Ibold_3]$ is the control input matrix.
Constraint~\eqref{eq:baseline_dynamics} enforces the dynamics via multiple
shooting. Constraint~\eqref{eq:baseline_control} bounds the impulse magnitude
from above by $u_{\max}$ and assigns a slack $\sigma_k$ to each maneuver
magnitude, which is minimized through~\eqref{eq:baseline_objective}.
In~\eqref{eq:baseline_geometry},
$\bm{E}=[\bm{I}_3\ \bm{0}_{3\times3}]$ extracts the position from the state,
$\bm{T}^{\mathrm{EM}}_{\mathrm{Inr}}(t_k)$ transforms that position from the
Moon-centered J2000 inertial frame to the Moon-centered Earth--Moon rotating
frame at $t_k$, and $\bm{c}_{s(k)}$ and $\bm{\Delta}_{s(k)}$ are,
respectively, the center and component-wise half-width vector of the geometry
box assigned to node $k$. The inequalities are applied componentwise. With the
quarter-period node spacing introduced above, the nodes in each revolution
correspond to the upper and lower out-of-plane extrema and the two intervening
central crossings. The mapping $s(k)$ assigns the upper extremum to the top
geometry box, the two crossings to the middle box, and the lower extremum to
the bottom box. The boxes preserve the broad vertical orbit topology without
enforcing pointwise agreement with the repeated CR3BP trajectory. Because the
baseline optimization is posed over a finite optimization window without
periodic endpoint closure, the first and last revolutions can exhibit boundary
effects; we therefore assess the geometry primarily over the interior
revolutions.

The nonlinear program is solved using SCvx*, a trust-region sequential convex programming method~\cite{Oguri2023SCvxStar} through SCPLib, with Clarabel~\cite{GoulartChen2026Clarabel} as the convex subproblem solver. Verner's most efficient 7/6 explicit Runge--Kutta method
\cite{Verner2010} propagates the nonlinear coast arcs and the associated
state-transition matrix (STM), with the HFEM Jacobian evaluated by
forward-mode automatic differentiation. The $L_2$ 1036 baseline-generation settings are summarized in Table~\ref{tab:baseline_setup}.

\begin{table}[t]
    \centering
    \caption{Baseline-generation setup for $L_2$ 1036}
    \label{tab:baseline_setup}
    \begin{tabular}{@{}p{0.34\textwidth}p{0.61\textwidth}@{}}
        \toprule
        Quantity & Setting \\
        \midrule
        Initial epoch & 2026 FEB 25 00:00:00 UTC \\
        CR3BP period / revolutions & $24.500847$ days / $82$ \\
        Nodes / coast arcs & $329$ / $328$ \\
        Duration & $2009.07$ days \\
        Frame & Moon-centered J2000 \\
        Perturbing bodies & Earth, Sun, Mercury, Venus, and Mars/Jupiter barycenters \\
        Lunar gravity & Degree/order $4$, Moon principal-axes frame \\
        Integrator & Verner $7/6$ Runge--Kutta,
        $\mathrm{reltol}=\mathrm{abstol}=10^{-14}$ \\
        Algorithm / convex solver & SCvx* / Clarabel \\
        Optimality / feasibility tol. & $10^{-6}$ / $10^{-10}$ \\
        Control bound & $u_{\max}=1$ (nondimensional; $700.20$~m/s equivalent) \\
        Maximum iterations & $500$ \\
        \bottomrule
    \end{tabular}
\end{table}

The HFEM/OCP variables use a numerical scaling separate from the Earth--Moon
CR3BP normalization above. The HFEM distance scale is
$\mathrm{DU}_{\mathrm{HF}}=10{,}000$~km, with corresponding velocity scale
$\mathrm{VU}_{\mathrm{HF}}=\sqrt{\mu_M/\mathrm{DU}_{\mathrm{HF}}}
=0.700200$~km/s. For the baseline-generation OCP, the nondimensional control
bound $u_{\max}=1$ therefore corresponds to $700.20$~m/s. This bound is a
numerical safeguard rather than a mission maneuver limit. Throughout this paper,
$\Delta V_{\mathrm{baseline}}
\equiv
\mathrm{VU}_{\mathrm{HF}}
\sum_{k=0}^{N_{\mathrm{node}}-1}\sigma_k$,
reported in m/s, denotes the dimensionalized baseline correction metric
obtained by scaling the optimized nondimensional objective by
$\mathrm{VU}_{\mathrm{HF}}$.

For baseline-generation sweeps that use same-date refinement, a
solution with $\Delta V_{\mathrm{baseline}} > 10^{-8}$ \SI{}{m/s} is automatically
re-solved at the same physical epoch using the solution as the new
initial guess. The $10^{-8}$~m/s threshold is used only to determine whether
an additional optimization solve is attempted. In this work, we describe a
reference as ballistic when the optimized deterministic correction is
sufficiently small to be operationally negligible over the reference duration.

For the start-epoch sweep, the 5.5-year $L_2$ 1036 baseline is generated at
90 consecutive daily start epochs. At each epoch requiring refinement, the
converged node states and all OCP control and scalar rows are reused as the next
initial guess while rebuilding a fresh problem and SCvx* algorithm at the same
physical epoch. The epoch, node times, force model, geometry boxes, and control
bounds remain unchanged, while the internal SCvx* state is reset.

\subsection{SPK Export and Validation}

Each optimized coast arc is exported as a separate Type 13 SPK segment
\cite{Acton1996,NAIFSPKRequiredReading}. The representative product uses
polynomial degree 7, an 1800-second sampling cadence, a $10^{-7}$-second gap
between adjacent segments, with Moon-centered state history in the J2000 frame.
The SPK is checked against the source trajectory and against short-arc propagation from the optimized
states. The SPK kernels generated for this study, together with the associated
validation software and metadata, are available in a versioned project
release.\footnote{Access release at:
\url{https://github.com/eXplorationLogisticsControl/spk-validations/tree/aas-sfm-2027-v1.0.0}}

\section{Station-Keeping and Dynamical Analysis}

This section formulates the station-keeping MPC and uncertainty model,
then defines the finite-time Lyapunov exponent and CR3BP stability-direction
analyses used to interpret the closed-loop behavior.

\subsection{Model Predictive Control}

The optimized SPK is treated as the nominal reference for the closed-loop problem.
Baseline generation is performed deterministically offline. Station keeping then
solves a finite-horizon OCP recursively from successive state estimates to reject
navigation, execution, and disturbance errors. At each control opportunity $t_i$, the state estimate is
\begin{equation}
    \hat{\bm{x}}_i
    =
    \bm{x}^{\mathrm{true}}_i+\bm{\eta}^{\mathrm{nav}}_i,
\end{equation}
where $\bm{\eta}^{\mathrm{nav}}_i$ is the six-component Cartesian navigation
error in position and velocity at control opportunity $i$.
We adopt the full-state targeting MPC
from~\cite{ShimaneEtAl2025OutputFeedback}, originally conceived for
Near-Rectilinear Halo Orbits (NRHO), where a finite-horizon impulsive problem is
solved at each control opportunity $i$. Let
$j=0,\ldots,N_{\mathrm{target}}$ index the prediction nodes, and let
$t_{j|i}$ denote their physical epochs with $t_{0|i}=t_i$. On coast arc $j$,
$\bm{x}(t)$ is initialized by
$\bm{x}(t_{j|i})=\bm{x}_{j|i}+\bm{B}\Delta\bm{v}_{j|i}$. The finite-horizon
problem is
\begin{subequations}\label{eq:mpc}
\begin{align}
    \min_{\bm{x}_{j|i},\,\Delta\bm{v}_{j|i},\,\sigma_{j|i}}\quad&
    \sum_{j=0}^{N_{\mathrm{target}}}\sigma_{j|i}
    \label{eq:mpc_objective}\\
    \mathrm{s.t.}\quad&
    \bm{x}_{0|i}=\hat{\bm{x}}_i,
    \label{eq:mpc_initial}\\
    &
    \bm{x}_{j+1|i}
    =
    \bm{x}_{j|i}+\bm{B}\Delta\bm{v}_{j|i}
    +\int_{t_{j|i}}^{t_{j+1|i}}
    \bm{f}\!\left(\bm{x}(t),t\right)\,\mathrm{d}t,
    \quad j=0,\ldots,N_{\mathrm{target}}-1,
    \label{eq:mpc_dynamics}\\
    &
    \|\Delta\bm{v}_{j|i}\|_2\leq\sigma_{j|i}\leq u_{\max},
    \quad j=0,\ldots,N_{\mathrm{target}},
    \label{eq:mpc_control}\\
    &
    \|\bm{r}_{N_{\mathrm{target}}|i}
      -\bm{r}^{\mathrm{ref}}_{N_{\mathrm{target}}|i}\|_2
    \leq\epsilon_r,
    \label{eq:mpc_position}\\
    &
    \|\bm{v}_{N_{\mathrm{target}}|i}
      +\Delta\bm{v}_{N_{\mathrm{target}}|i}
      -\bm{v}^{\mathrm{ref}}_{N_{\mathrm{target}}|i}\|_2
    \leq\epsilon_v.
    \label{eq:mpc_velocity}
\end{align}
\end{subequations}
Here $\bm{x}_{j|i}$, $\Delta\bm{v}_{j|i}$, and $\sigma_{j|i}$ are the predicted
pre-impulse state, planned impulse, and slack variable, respectively.
Constraint~\eqref{eq:mpc_initial} initializes each MPC problem with the current
navigation estimate $\hat{\bm{x}}_i$.
Constraints~\eqref{eq:mpc_position} and~\eqref{eq:mpc_velocity} bound the
terminal position and post-impulse velocity to lie within an ellipsoid with
radii $\epsilon_r$ and $\epsilon_v$.
The superscript ``ref'' denotes the SPK reference at
$t_{N_{\mathrm{target}}|i}$. Only the first planned impulse is implemented on
the spacecraft, after which the problem is rebuilt from a new navigation
estimate at the next control opportunity $t_{i+1}$.

For $L_2$ 1036, the station-keeping schedule uses two control opportunities
per CR3BP revolution, located at the upper and lower out-of-plane amplitude
extrema, referred to here as the top- and bottom-side opportunities,
respectively. Consecutive control opportunities are therefore separated by
approximately half a revolution, so a prediction horizon of
$N_{\mathrm{target}}$ control intervals spans
$N_{\mathrm{target}}/2$ revolutions.

\subsection{Uncertainty and Disturbance Models}

Initial insertion and navigation errors are modeled as independent, zero-mean
Gaussian errors in each Cartesian state component. The uncertainty and
disturbance parameters used in the station-keeping simulations are summarized
in Table~\ref{tab:uncertainties}. The low and high navigation error levels and the maneuver-execution and SRP
models follow the scale and structure of prior high-fidelity NRHO maintenance
studies
\cite{DavisEtAl2017,DavisEtAl2022,KayserEtAl2022,ThompsonEtAl2021}.
The attitude-deadband and attitude-slew errors follow Davis et
al.~\cite{DavisEtAl2017}, where mismodeled attitude-control activity is
represented by fixed-magnitude impulses in random directions.

\begin{table}[t]
    \centering
    \caption{Uncertainty and disturbance assumptions}
    \label{tab:uncertainties}
    \begin{threeparttable}
    \centering
    \renewcommand{\arraystretch}{1.08}
    \begin{tabular}{@{}p{0.38\textwidth}p{0.57\textwidth}@{}}
        \toprule
        Source & Assumption \\
        \midrule
        Initial insertion error &
        Position $150$~km; velocity $50$~cm/s (component-wise $3\sigma$) \\
        \addlinespace[0.15em]

        Low navigation error &
        Position $1$~km; velocity $1$~cm/s (component-wise $3\sigma$) \\

        High navigation error &
        Position $10$~km; velocity $10$~cm/s (component-wise $3\sigma$) \\
        \addlinespace[0.15em]

        Maneuver magnitude error &
        $3\%$ multiplicative ($3\sigma$) \\

        Additive maneuver execution error &
        $0.03$~cm/s in a random direction \\

        $C_R$ uncertainty &
        $30\%$ fractional ($3\sigma$) per control interval \\

        $A/m$ uncertainty &
        $15\%$ fractional ($3\sigma$) per control interval \\

        Attitude-deadband error &
        $2.0043\times10^{-5}$~m/s in a random direction every $70$~min \\
        \addlinespace[0.15em]

        Attitude-slew error &
        $6.9751\times10^{-4}$~m/s in a random direction every $3.2$~hr \\
        \bottomrule
    \end{tabular}
    \begin{tablenotes}[flushleft]
        \footnotesize
        \item Navigation, maneuver-execution, and SRP assumptions follow the scale and
structure of prior high-fidelity NRHO maintenance studies
\cite{DavisEtAl2017,DavisEtAl2022,KayserEtAl2022,ThompsonEtAl2021}.
The attitude-deadband and attitude-slew errors follow
Davis et al.~\cite{DavisEtAl2017}; the initial insertion error is
study-specific.
    \end{tablenotes}
    \end{threeparttable}
\end{table}

Maneuvers below 0.15~cm/s are not executed. The minimum-maneuver threshold,
maneuver-execution errors, and adopted spacecraft-disturbance structure follow
previous works on NRHO maintenance~\cite{DavisEtAl2017,DavisEtAl2022}.

OCP arcs use Verner's most efficient 7/6 explicit Runge--Kutta method, and truth
arcs use the corresponding most efficient 8/7 method~\cite{Verner2010}. The OCP
relative and absolute tolerances are $10^{-12}$, while the truth relative and
absolute tolerances are $10^{-12}$ and $10^{-14}$, respectively. SCvx* uses at most 300 iterations for the $L_2$ 1036 station-keeping OCPs,
with optimality and feasibility tolerances of $10^{-3}$ and $10^{-6}$,
respectively. The control
bound $u_{\max}=10$, corresponding to $7002$~m/s under the HFEM/OCP scaling,
provides a loose numerical upper bound on the magnitude of each planned impulse.

The principal closed-loop tracking quantities are the position and velocity
deviation magnitudes
\begin{equation}
    \Delta r(t)=\|\bm{r}^{\mathrm{true}}(t)-\bm{r}^{\mathrm{ref}}(t)\|_2,
    \qquad
    \Delta v(t)=\|\bm{v}^{\mathrm{true}}(t)-\bm{v}^{\mathrm{ref}}(t)\|_2,
    \label{eq:tracking_deviation}
\end{equation}
together with executed cumulative $\Delta V$ and its annualized value. A
realization is counted as successful when the full simulation horizon
is completed without loss of a feasible solution to~\eqref{eq:mpc}.

\subsection{Finite-Time Lyapunov Exponent}

The finite-time Lyapunov exponent (FTLE) characterizes local amplification over
a finite interval along the high-fidelity reference
\cite{ShaddenLekienMarsden2005}. For each selected start phase, the HFEM STM is
propagated over one CR3BP period $T_{\mathrm{C}}$, consistent with related
one-revolution NRHO analyses~\cite{ShimaneEtAl2025ACC}. Before computing
singular values, the state is scaled with
$\bm{D}=\operatorname{diag}(r_s\bm{I}_3,v_s\bm{I}_3)$, where
$r_s=10{,}000$~km and $v_s=0.700200$~km/s. The scaled STM is
$\widetilde{\bm{\Phi}}=\bm{D}^{-1}\bm{\Phi}\bm{D}$, and we define
\begin{equation}
    \Lambda(t)
    =
    \ln \sigma_{\max}\!\left(
    \widetilde{\bm{\Phi}}(t+T_{\mathrm{C}},t)
    \right),
    \qquad
    \lambda_{\mathrm{FTLE}}(t)
    =
    \frac{\Lambda(t)}{T_{\mathrm{C}}}.
    \label{eq:ftle}
\end{equation}
Here, $\sigma_{\max}$ is the largest singular value of the scaled STM,
$\lambda_{\mathrm{FTLE}}$ is the one-period FTLE, and $\Lambda$ is the
corresponding unnormalized logarithmic stretch.

\subsection{CR3BP Stability Directions and Maneuver Angles}

The underlying periodic CR3BP orbit is used separately to define local comparison
directions for the planned maneuvers. For $L_2$ 1036, the one-period CR3BP STM
defines the monodromy matrix
\begin{equation}
    \bm{M}_{\mathrm{C}}=\bm{\Phi}_{\mathrm{C}}(T_{\mathrm{C}},0).
\end{equation}
The eigenvectors associated with the smallest- and largest-magnitude
eigenvalues define the stable and unstable axes at the initial CR3BP phase,
respectively, with the corresponding eigenvalues satisfying
$|\lambda_s|<1$ and $|\lambda_u|>1$
\cite{GomezMondelo2001,ShirobokovEtAl2017}. We denote these eigenvectors
$\bm{\nu}_{s,0}$ and $\bm{\nu}_{u,0}$.

At each station-keeping opportunity, we identify the corresponding phase of
the underlying periodic CR3BP orbit from the elapsed time since the reference
initial epoch. The corresponding within-revolution time is denoted by
$\tau_i$. The two eigendirections are transported from the initial CR3BP phase
to $\tau_i$ with the CR3BP STM,
\begin{equation}
    \bm{\nu}_{s,i}
    =
    \bm{\Phi}_{\mathrm{C}}(\tau_i,0)\bm{\nu}_{s,0},
    \qquad
    \bm{\nu}_{u,i}
    =
    \bm{\Phi}_{\mathrm{C}}(\tau_i,0)\bm{\nu}_{u,0}.
    \label{eq:phase_transport_eig}
\end{equation}
Because the MPC control is an impulsive velocity change, the comparison uses
the velocity components of these transported six-dimensional directions. With
$\bm{E}_v=[\bm{0}_{3\times3}\ \bm{I}_3]$,
\begin{equation}
    \hat{\bm e}_{s,i}
    =
    \frac{\bm E_v\bm{\nu}_{s,i}}
    {\|\bm E_v\bm{\nu}_{s,i}\|_2},
    \qquad
    \hat{\bm e}_{u,i}
    =
    \frac{\bm E_v\bm{\nu}_{u,i}}
    {\|\bm E_v\bm{\nu}_{u,i}\|_2}.
\end{equation}
These transported CR3BP directions serve as local comparison axes for
interpreting the planned maneuver direction. The planned first MPC maneuver is expressed in J2000 and rotated into the Moon-centered Earth--Moon rotating frame with the three-dimensional frame
rotation at the maneuver epoch. We normalize the planned maneuver as
$\Delta\hat{\bm v}_i=\Delta\bm v_i/\|\Delta\bm v_i\|_2$ and compute the
acute axis angles
\begin{equation}
    \theta_{s,i}=\cos^{-1}\!\left(\left|
    \Delta\hat{\bm v}_i^T\hat{\bm e}_{s,i}\right|\right),
    \qquad
    \theta_{u,i}=\cos^{-1}\!\left(\left|
    \Delta\hat{\bm v}_i^T\hat{\bm e}_{u,i}\right|\right).
    \label{eq:stability_axis_angles}
\end{equation}
The absolute value treats $+\bm e$ and $-\bm e$ as the same one-dimensional
axis, consistent with eigenvector sign ambiguity. We include only planned first
maneuvers with $\|\Delta\bm v\|_2\geq0.15$~cm/s, matching the execution threshold,
and evaluate top and bottom opportunities separately for the low and high
navigation error cases. Executed or execution-corrupted maneuvers are not used in this
comparison.

\section{Numerical Results}

We first examine in detail the $L_2$ 1036 reference and its closed-loop
maintenance. The baseline-generation analysis is then extended to start-epoch
robustness and the 200-case $L_1/L_2$ family. Station-keeping performance is
subsequently compared across selected long-duration $L_1$ and $L_2$
vertical-orbit references.

\subsection{\texorpdfstring{Detailed $L_2$ 1036 Analysis}
{Detailed L2 1036 Analysis}}

\subsubsection{Baseline Generation and SPK Validation}

\begin{figure}[!htbp]
    \centering
    \includegraphics[width=0.90\textwidth]{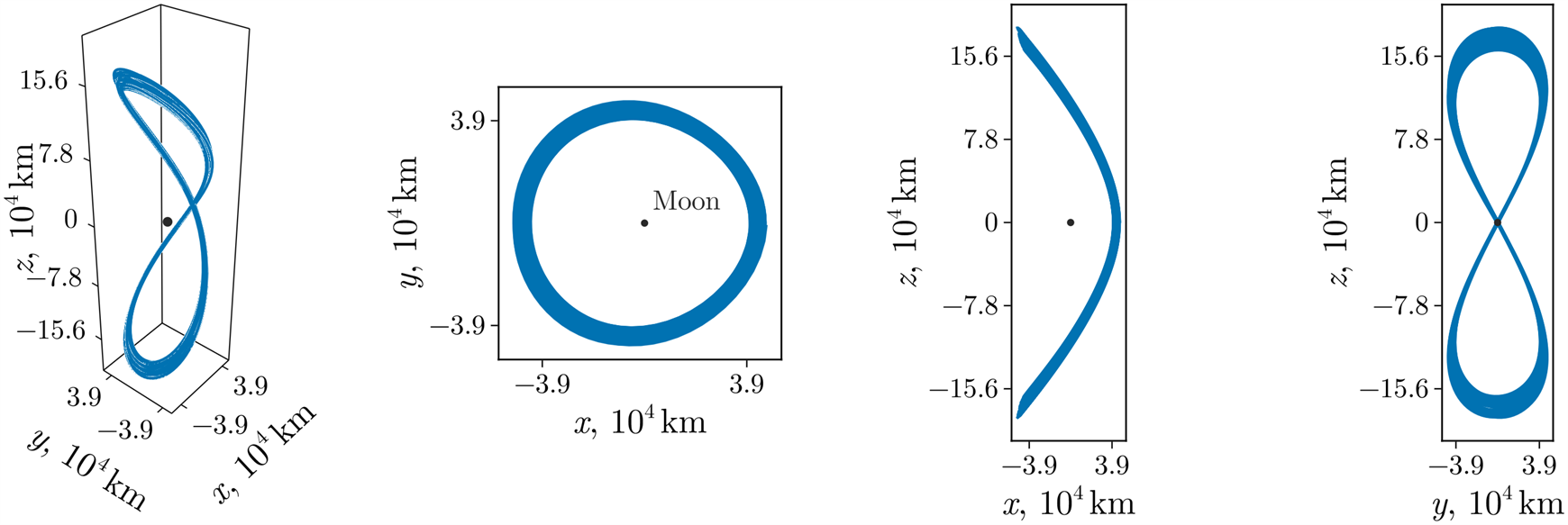}
    \caption{5.5-year $L_2$ 1036 high-fidelity reference in the
    Moon-centered Earth--Moon rotating frame}
    \label{fig:primary_baseline}
\end{figure}

The $L_2$ 1036 solve converges to a locally optimal solution in 31 SCvx*
iterations without a warm-start refinement. The optimized baseline correction is $\Delta V_{\mathrm{baseline}}=1.16\times10^{-10}$~m/s, and the largest scalar
node control is $6.31\times10^{-13}$~m/s. The maximum nonlinear constraint
violation is $7.96\times10^{-11}$ under the high-fidelity optimal-control
scaling, corresponding to approximately $7.96\times10^{-4}$~m and
$5.57\times10^{-8}$~m/s for position and velocity components, respectively.
The optimized deterministic correction is negligible at an operational scale,
so we refer to this 5.5-year reference as ballistic.
Figure~\ref{fig:primary_baseline} shows the resulting geometry; for
visualization, the plotted interval ends at the last complete upward $z=0$
crossing before the SPK endpoint. Table~\ref{tab:representative_baseline}
summarizes the solve.

\begin{table}[t]
    \centering
    \caption{$L_2$ 1036 baseline result.}
    \label{tab:representative_baseline}
    \begin{tabular}{ll}
        \toprule
        Quantity & Result \\
        \midrule
        SCvx* iterations & $31$ \\
        Correction cost & $1.16\times10^{-10}$~m/s \\
        Maximum node control & $6.31\times10^{-13}$~m/s \\
        Maximum constraint violation & $7.96\times10^{-11}$ \\
        Revolutions / nodes / arcs & $82$ / $329$ / $328$ \\
        Duration & $2009.07$~days \\
        Top / middle / bottom nodes & $82$ / $165$ / $82$ \\
        Warm-start refinement & None \\
        \bottomrule
    \end{tabular}
\end{table}

The optimized reference contains 328 coast arcs separated by impulsive nodes.
A separate Type 13 SPK segment is generated for each coast arc, preserving the
velocity discontinuities at the maneuver nodes. To verify the SPK
interpolation, states evaluated from the kernel are compared directly with the
corresponding optimized reference states, giving position residuals of order
$10^{-10}$~km and peak velocity residuals of order $10^{-12}$~km/s. As a
dynamical consistency check, states obtained from the SPK are also propagated
with the high-fidelity model over short node-to-node intervals and compared
with the SPK state at the interval endpoint. The resulting position
differences remain below about $10^{-7}$~km, while the velocity differences
remain of order $10^{-12}$~km/s. We use the validated SPK as the reference
trajectory for the subsequent station-keeping analyses.

\subsubsection{Geometry Constraints and Start-Epoch Robustness}

We first test the effect of the geometry constraints by uniformly scaling the
component-wise half-width vectors $\bm{\Delta}_{s(k)}$ of the top, middle, and
bottom rotating-frame geometry boxes defined in
Eq.~\eqref{eq:baseline_geometry}. For this sweep, all three box types use
half-widths of
$s_{\mathrm{box}}[2,2,2]^T\times10^5$~km, where $s_{\mathrm{box}}$ is the
common box-width scale factor. We test 40 box scales
spanning $s_{\mathrm{box}}=0.0025$ to $2.0$. Apsis geometry is measured by the absolute difference in Moon-centered scalar
radius between matched high-fidelity and repeated-CR3BP apsis events. For
matched apsis event $j$, let $\bm{r}^{\mathrm{HF}}_j$ and
$\bm{r}^{\mathrm{C}}_j$ denote the corresponding Moon-centered position
vectors on the high-fidelity and repeated-CR3BP trajectories, respectively.
The apsis-radius difference is
$|\Delta r_{\mathrm{apsis},j}|
=
\left|
\|\bm{r}^{\mathrm{HF}}_j\|_2
-
\|\bm{r}^{\mathrm{C}}_j\|_2
\right|$.
A node is classified as active when at least one component of its
rotating-frame position lies on the corresponding geometry-box boundary. At the tightest tested scale, $s_{\mathrm{box}}=0.0025$, every node is active
under the box-face diagnostic and the baseline requires
$12.497$~\SI{}{km/s} of correction. The median apolune and
perilune radius differences are about $607$~\SI{}{km} and $280$~\SI{}{km}. As the boxes are widened
in this $L_2$ 1036 sweep, the optimizer receives more geometric freedom, and the
required correction drops sharply through the transition region. We treat this
as an observed sensitivity for this case only; it does not establish a global
or monotonic family-wide relation between box width, geometry, and correction
cost.

The main transition occurs between the tested scales $s_{\mathrm{box}}=0.14$
and $0.15$. At $s_{\mathrm{box}}=0.14$, the cost is $8.318$~m/s and $2.13\%$ of the
nodes remain active. At $s_{\mathrm{box}}=0.15$, no nodes are active and the
cost is $6.17\times10^{-10}$~m/s. Thus $s_{\mathrm{box}}=0.15$ is the smallest
tested inactive box for this specific case. At the production scale
$s_{\mathrm{box}}=1$, the median apolune and perilune radius offsets are about
$13{,}068$ and $1{,}433$~km and the correction is $1.16\times10^{-10}$~m/s.

\begin{figure}[]
    \centering
    \includegraphics[width=0.90\textwidth]{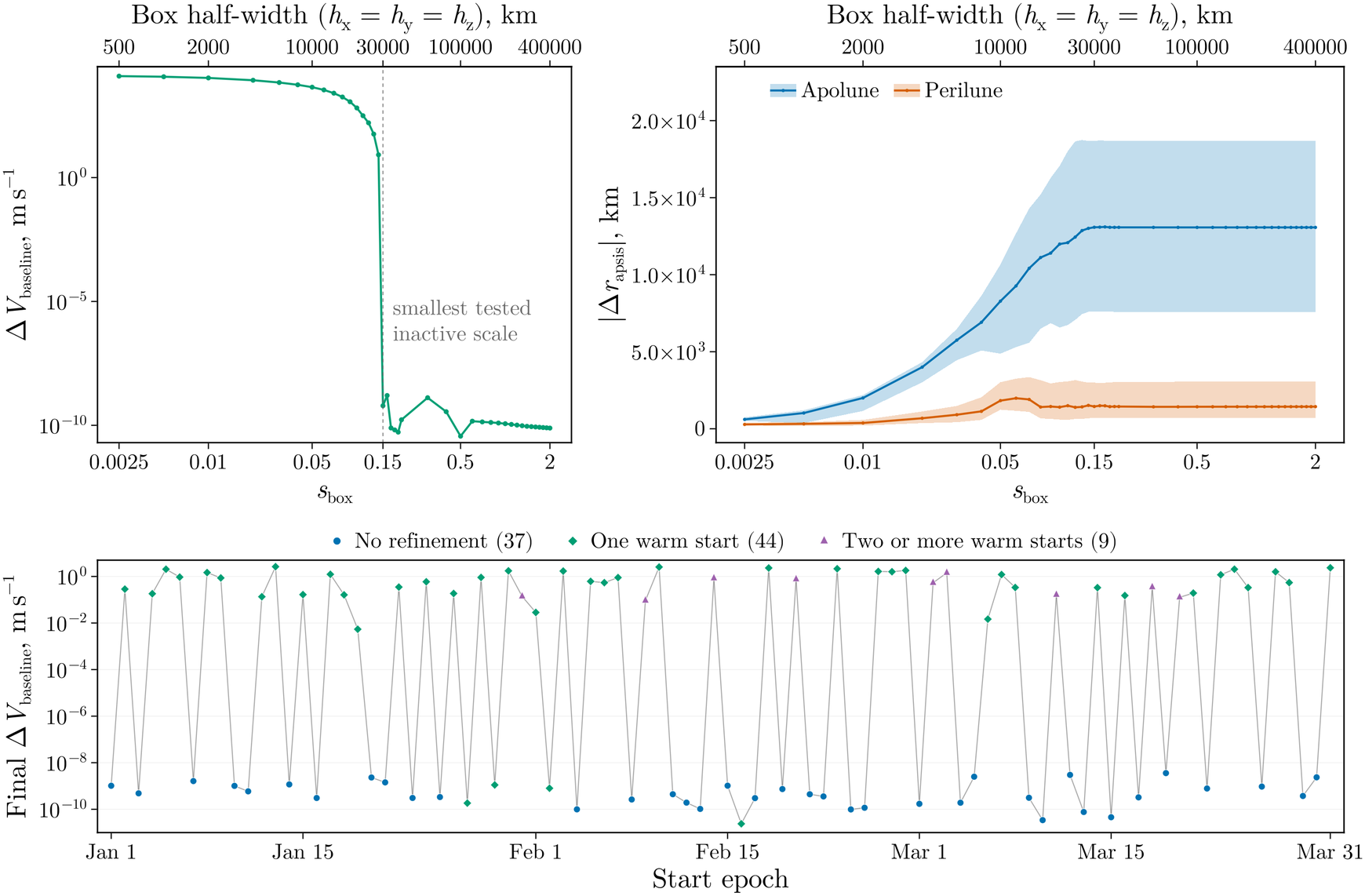}
    \caption{$L_2$ 1036 geometry-constraint sensitivity and start-epoch
robustness across 90 daily epochs}
    \label{fig:baseline_sensitivity}
\end{figure}

We then test start-epoch robustness by solving the same 5.5-year $L_2$ 1036
problem at 90 consecutive daily epochs from 1 January through 31 March 2026.
After same-date refinement where needed, a locally optimal solution of the
baseline-generation OCP is obtained at each of the 90 tested start epochs.
Across the 90 start epochs, 37 require no refinement, 44 require one warm
start, and 9 require two or more warm starts. Every recorded refinement pass
reduces $\Delta V_{\mathrm{baseline}}$.
Figure~\ref{fig:baseline_sensitivity} shows the box sweep together with the
final start-epoch results.

The lower panel of Fig.~\ref{fig:baseline_sensitivity} summarizes the
start-epoch sweep. Because the number of same-date refinement passes varies
among the 90 start epochs, the final $\Delta V_{\mathrm{baseline}}$ values
depend on the refinement history as well as the selected epoch. We therefore
use this sweep to assess whether the baseline-generation procedure can obtain
a locally optimal solution across the tested start dates; the final correction
levels are not used to infer a physical dependence of baseline cost on epoch.

\subsubsection{Finite-Time Lyapunov Exponent}

\begin{figure}[]
    \centering
    \includegraphics[width=0.90\textwidth]{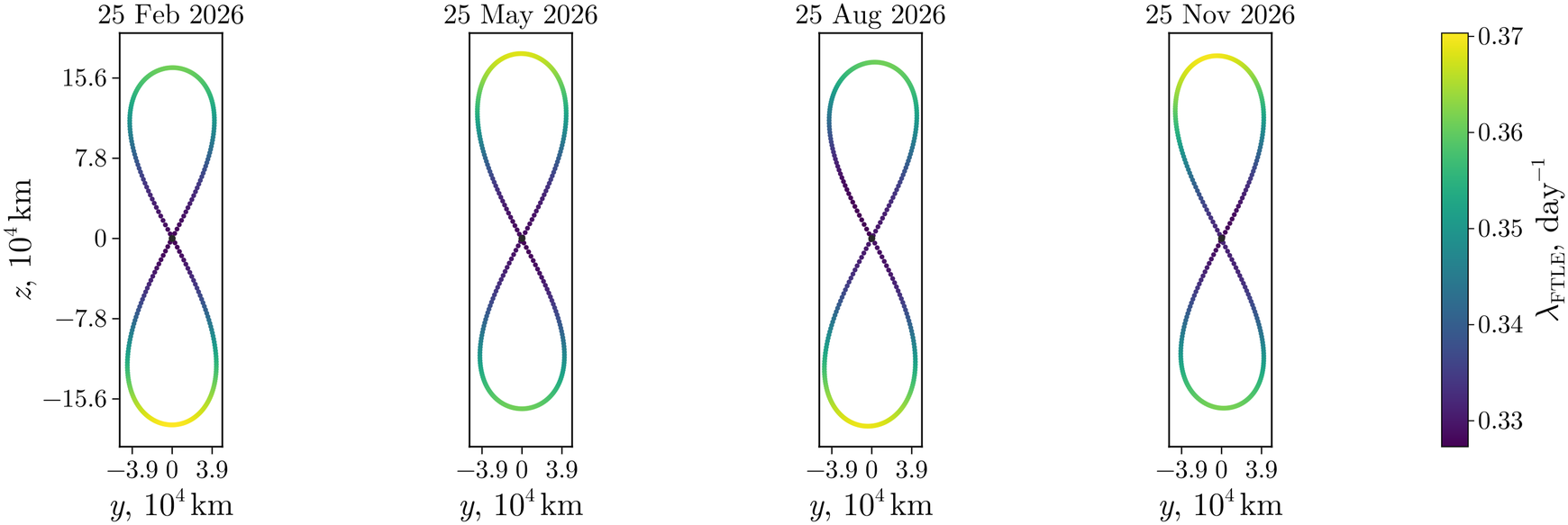}
    \caption{One-revolution finite-time Lyapunov exponent,
$\lambda_{\mathrm{FTLE}}$, along the $L_2$ 1036 high-fidelity reference over
300 start phases at each of four epochs}
    \label{fig:ftle}
\end{figure}

We evaluate the one-period FTLE over 300 start phases spanning one CR3BP
revolution at each of four epochs: 25 February, 25 May, 25 August, and
25 November 2026. For each sampled phase, the HFEM STM is propagated over one
CR3BP period, $T_{\mathrm{C}}=24.500847$~days. Across the 1200 samples,
$\Lambda=\ln\sigma_{\max}$ ranges from $8.019$ to $9.074$, corresponding to
$\lambda_{\mathrm{FTLE}}=0.3273$--$0.3703$~day$^{-1}$. The lowest FTLE remains near the central crossing, while the location of the
largest FTLE shifts between the upper and lower large-$|z|$ portions of the
orbit across the four epochs. This epoch dependence is consistent with the
time-dependent HFEM dynamics and likely reflects changes in the
Sun--Earth--Moon geometry, including the changing orientation of the
Earth--Moon orbital plane relative to the Sun--Earth direction. Figure~\ref{fig:ftle} shows the FTLE distribution along the orbit at each
epoch.

\subsubsection{Controller Selection and Disturbance Effects}

\begin{figure}[]
    \centering

    \includegraphics[width=0.666\textwidth]{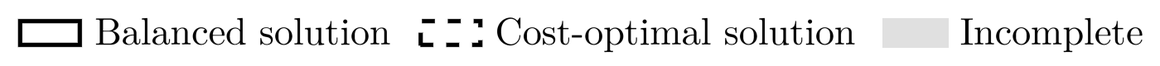}

    \vspace{0.2em}

    \begin{subfigure}[t]{0.32\textwidth}
        \centering
        \includegraphics[width=0.90\linewidth]{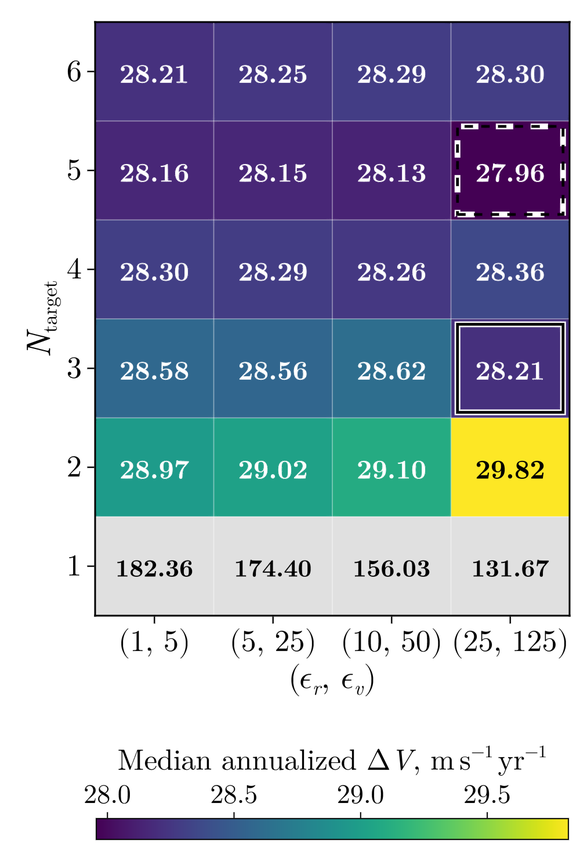}
        \captionsetup{
            justification=centering,
            singlelinecheck=false,
            margin={0.06\linewidth,0pt}
        }
        \caption{Maneuver-cost\\ sensitivity}
        \label{fig:controller_selection_cost}
    \end{subfigure}\hfill
    \begin{subfigure}[t]{0.32\textwidth}
        \centering
        \includegraphics[width=0.90\linewidth]{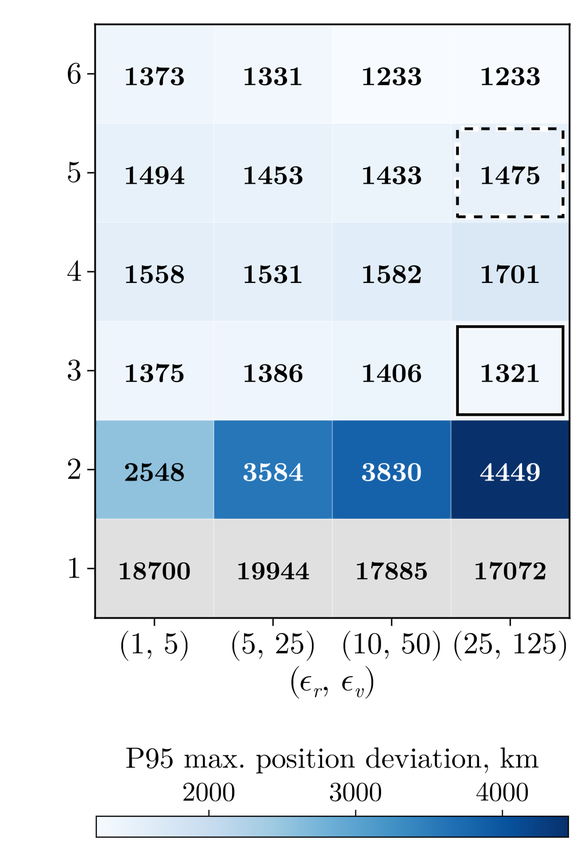}
        \captionsetup{
            justification=centering,
            singlelinecheck=false,
            margin={0.06\linewidth,0pt}
        }
        \caption{Position-tracking\\ sensitivity}
        \label{fig:controller_selection_position}
    \end{subfigure}\hfill
    \begin{subfigure}[t]{0.32\textwidth}
        \centering
        \includegraphics[width=0.90\linewidth]{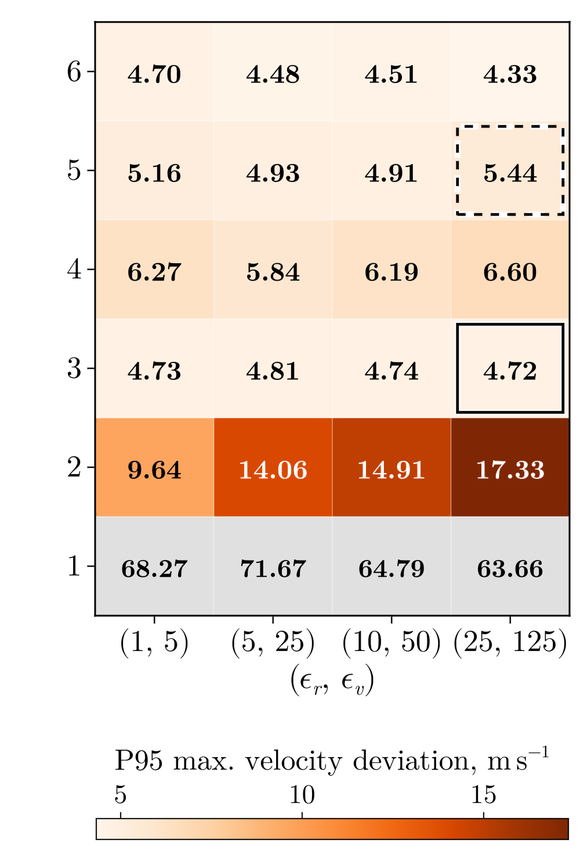}
        \captionsetup{
            justification=centering,
            singlelinecheck=false,
            margin={0.06\linewidth,0pt}
        }
        \caption{Velocity-tracking\\ sensitivity}
        \label{fig:controller_selection_velocity}
    \end{subfigure}

    \caption{Controller performance sensitivity to targeting horizon and terminal constraint tolerances}
    \label{fig:controller_selection}
\end{figure}

For the controller-selection study, we test six horizons,
$N_{\mathrm{target}}\in\{1,2,3,4,5,6\}$. The terminal position and velocity
tolerances are denoted by $(\epsilon_r,\epsilon_v)$. We test four pairs:
$(1~\mathrm{km},5~\mathrm{m/s})$, $(5~\mathrm{km},25~\mathrm{m/s})$,
$(10~\mathrm{km},50~\mathrm{m/s})$, and
$(25~\mathrm{km},125~\mathrm{m/s})$, giving 24 controller configurations.
Each configuration is evaluated with 100 one-year Monte Carlo realizations
under low navigation error and 100 under high navigation error, giving
4800 realizations across the full controller-selection study.
Figure~\ref{fig:controller_selection} shows the trade between maneuver cost
and 95th-percentile maximum position and velocity deviations under high
navigation error.

Among the tested configurations, $N_{\mathrm{target}}=5$ with
$(\epsilon_r,\epsilon_v)=(25~\mathrm{km},125~\mathrm{m/s})$ gives the lowest
median annualized maneuver cost. The configuration with
$N_{\mathrm{target}}=3$ and the same terminal tolerances gives nearly the same
cost while reducing the 95th-percentile maximum position and velocity
deviations. Under high navigation error, its median annualized cost is
$28.21$~m/s/yr compared with $27.96$~m/s/yr for
$N_{\mathrm{target}}=5$, an increase of only $0.905\%$, while
the 95th-percentile maximum position and velocity deviations decrease from
$1475$ to $1321$~km and from $5.44$ to $4.72$~m/s, respectively. We assume the MPC is
solved on the ground, so solve time is not a primary comparison metric.
Based on this cost--tracking trade, we carry forward
$N_{\mathrm{target}}=3$ and
$(\epsilon_r,\epsilon_v)=(25~\mathrm{km},125~\mathrm{m/s})$ as the balanced
comparison configuration for the subsequent studies. Table~\ref{tab:controller_selection} compares this configuration with the cost-optimal $N_{\mathrm{target}}=5$ case.

\begin{table}[t]
    \centering
    \caption{Performances with balanced and cost-optimal controller configurations}
    \label{tab:controller_selection}
    \setlength{\tabcolsep}{5pt}
    \begin{tabular}{@{}p{0.53\textwidth}cc@{}}
        \toprule
        Configuration: $(\epsilon_r,\epsilon_v)$ [\SI{}{km}, \SI{}{m/s}] & Balanced: $(25,125)$ & Cost-optimal: $(25,125)$ \\
        \midrule
        MPC prediction horizon, $N_{\mathrm{target}}$
        & $3$ & $5$ \\

        High navigation error median $\Delta V$, m/s/yr
        & $28.21$ & $27.96$ \\

        High navigation error 95th-percentile $\Delta V$, m/s/yr
        & $36.54$ & $36.53$ \\

        High navigation error 95th-percentile max\ $\Delta r$, km
        & $1321$ & $1475$ \\

        High navigation error 95th-percentile max\ $\Delta v$, m/s
        & $4.72$ & $5.44$ \\
        \bottomrule
    \end{tabular}
\end{table}

All ten navigation/disturbance combinations complete successfully. As shown in
Table~\ref{tab:disturbance_decomposition}, at the levels tested here,
navigation uncertainty changes the median annualized cost more than any
isolated SRP or attitude disturbance. The closed-loop propagation and repeated
MPC solves are nonlinear, so the combined disturbance response is not
necessarily equal to the sum of the responses from the isolated cases.

The five disturbance cases vary the SRP-parameter uncertainty and
attitude-disturbance errors while holding the remaining Monte Carlo assumptions
fixed for a given navigation-error level. The quiet case uses nominal SRP with
fixed $C_R$ and $A/m$ and omits the attitude-deadband and attitude-slew errors.
The three isolated cases include only the named SRP-parameter uncertainty,
attitude-deadband error, or attitude-slew error. The all-disturbances case
includes all three.
\begin{table}[]
    \centering
    \caption{Median annualized station-keeping cost in the disturbance decomposition}
    \label{tab:disturbance_decomposition}
    \begin{tabular}{lcc}
        \toprule
        Disturbance case &
        \multicolumn{2}{c}{$\Delta V$, m/s/yr} \\
        & \shortstack{Low navigation error} &
          \shortstack{High navigation error} \\
        \midrule
        Quiet & $3.22$ & $27.98$ \\
        SRP uncertainty only & $3.35$ & $28.66$ \\
        Attitude-deadband error only & $3.13$ & $28.33$ \\
        Attitude-slew error only & $3.42$ & $29.07$ \\
        All disturbances & $3.74$ & $28.21$ \\
        \bottomrule
    \end{tabular}
\end{table}

\FloatBarrier

\subsubsection{Five-Year Station Keeping}

\begin{figure}[]
    \centering
    \includegraphics[width=0.80\textwidth]{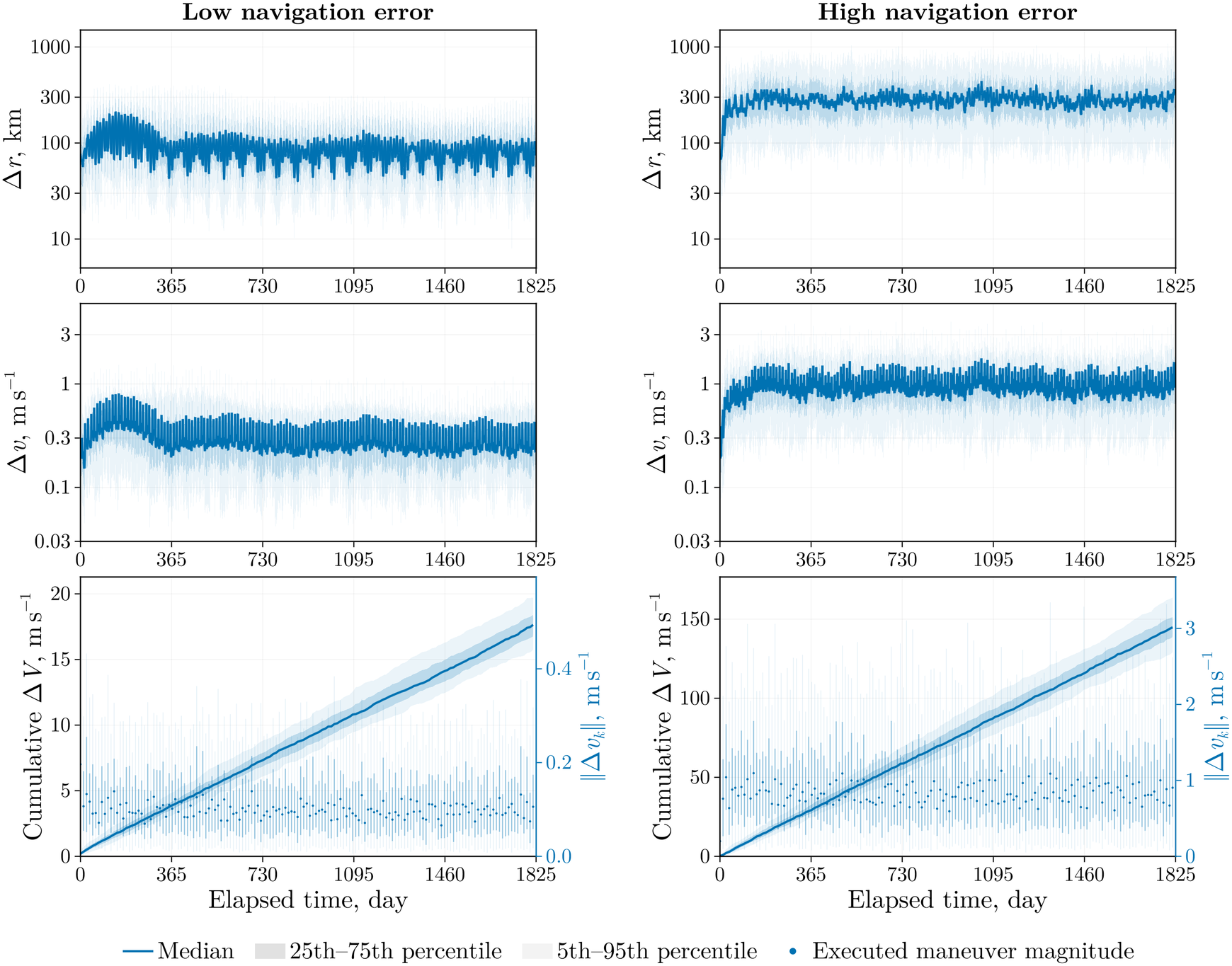}
    \caption{Five-year Monte Carlo station-keeping histories for the balanced comparison configuration under low and high navigation error}
    \label{fig:five_year}
\end{figure}

We consider a long-duration test with the balanced comparison configuration
$N_{\mathrm{target}}=3$ and
$(\epsilon_r,\epsilon_v)=(25~\mathrm{km},125~\mathrm{m/s})$.
We propagate 100 realizations each at low and high navigation errors for
five years. All 200 runs complete
successfully, where each run contains 149 receding-horizon control
opportunities. Figure~\ref{fig:five_year} shows the ensemble histories: curves
are medians, dark and light bands denote the 25th--75th and 5th--95th
percentiles, respectively, and the bottom-row maneuver markers use the same
median/interquartile/5th--95th-percentile convention with sub-threshold
maneuvers that are not executed included as zero. Table~\ref{tab:five_year_mc} summarizes the cost and maximum
tracking deviations.

\begin{table}[t]
    \centering
    \caption{Five-year Monte Carlo station-keeping performance}
    \label{tab:five_year_mc}
    \setlength{\tabcolsep}{4pt}
    \begin{tabular}{@{}lrrrr@{}}
        \toprule
        & \multicolumn{2}{c}{Low navigation error}
        & \multicolumn{2}{c}{High navigation error} \\
        \cmidrule(lr){2-3}\cmidrule(lr){4-5}
        Metric & Median & 95th percentile & Median & 95th percentile \\
        \midrule
        Total executed $\Delta V$, m/s
        & $17.643$ & $19.71$ & $144.860$ & $163.59$ \\

        Annualized executed $\Delta V$, m/s/yr
        & $3.530$ & $3.94$ & $28.987$ & $32.74$ \\

        Maximum position deviation, km
        & $315.3$ & $470$ & $1127.2$ & $1687$ \\

        Maximum velocity deviation, m/s
        & $1.226$ & $1.90$ & $3.838$ & $6.86$ \\
        \bottomrule
    \end{tabular}
\end{table}

The navigation-error level has a clear effect on both cost and tracking. Relative
to the low navigation error case, the high navigation error case increases the
median annualized cost by a factor of $8.21$ and the median maximum position
deviation by a factor of $3.57$. In the low navigation error case, the total executed cost ranges from $14.63$ to $21.03$~m/s,
with a mean of $17.61$~m/s and a median of $17.64$~m/s. In the high navigation error case, the range is $125.81$--$170.34$~m/s,
with a mean of $145.72$~m/s and a median of $144.86$~m/s. The close mean and median values show
that these results are not being set by only a few extreme runs.

The larger maneuver cost is accompanied by larger tracking dispersion: the
median root-mean-square position deviation increases from $109.4$ to $348.6$~km. The fixed-size MPC problems are readily solved in the current ground-based
implementation. For the high navigation error runs, the median of the per-run
mean solve times is $4.03$~s per OCP and the 95th percentile is $4.29$~s.
Because cislunar orbit-maintenance maneuvers are typically planned on the
ground and uploaded to the spacecraft, onboard real-time solution is not
required here.

\begin{figure}[]
    \centering
    \includegraphics[width=0.80\textwidth]{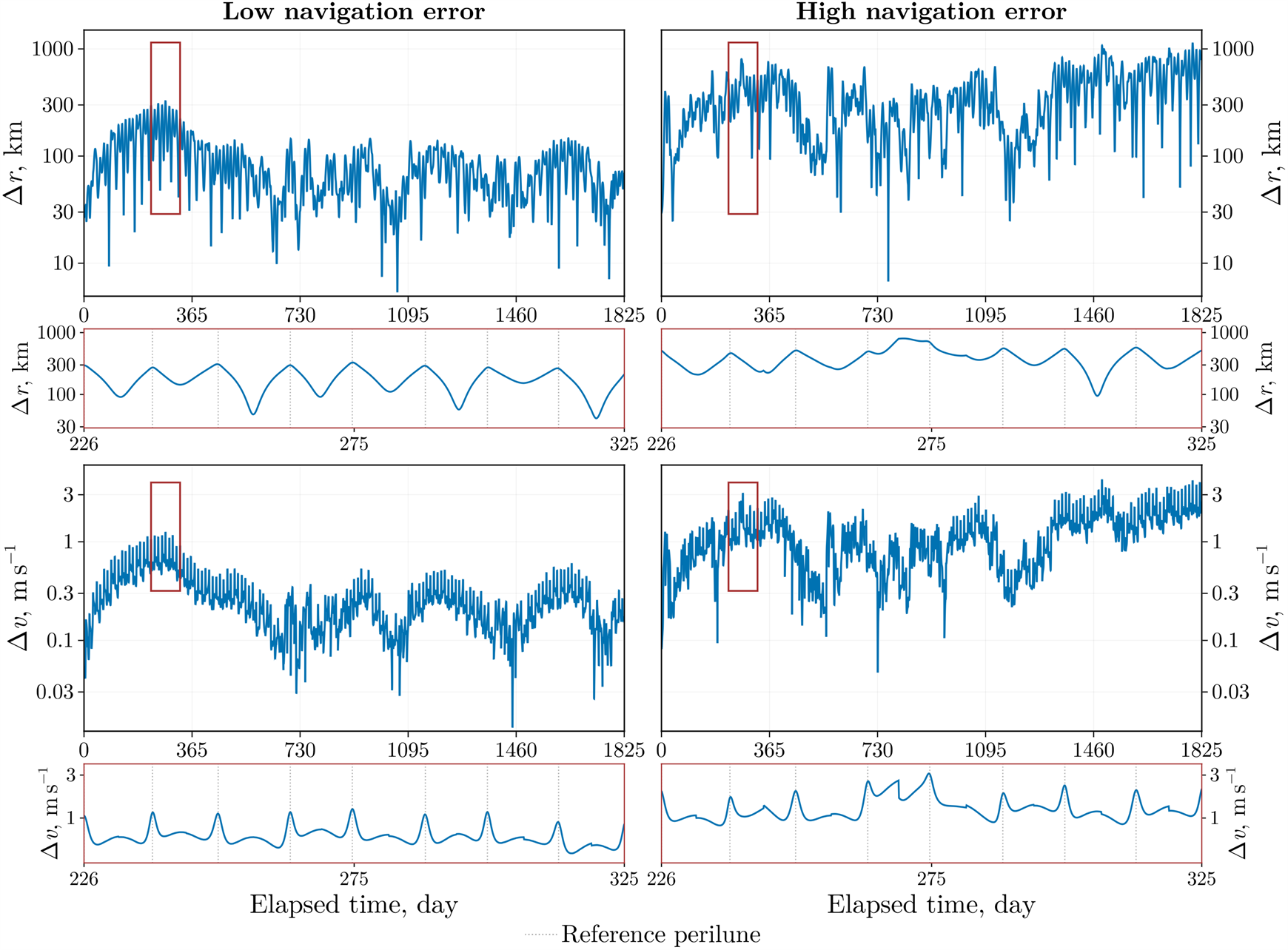}
    \caption{Representative deviation histories for the balanced comparison configuration, with common zoom intervals and reference perilunes marked by dotted lines}
    \label{fig:representative_deviation_history}
\end{figure}

Figure~\ref{fig:representative_deviation_history} complements the ensemble
statistics in Fig.~\ref{fig:five_year} by showing a representative realization selected independently from each navigation-error ensemble.
The full histories show the position and velocity deviation magnitudes defined
in Eq.~\eqref{eq:tracking_deviation}, while the attached strips show the same
common multi-revolution interval in greater detail.

\subsubsection{Maneuver Alignment with CR3BP Stability Directions}

The FTLE characterizes finite-time amplification along the high-fidelity
reference. The CR3BP stable and unstable axes are used separately to interpret
the directions of the planned MPC maneuvers. For $L_2$ 1036, the stable- and unstable-eigenvalue magnitudes of the CR3BP
monodromy matrix are $2.65354\times10^{-3}$ and
$3.76855\times10^{2}$, respectively. We use the corresponding
phase-transported velocity directions only as local comparison axes for the
planned maneuvers.

\begin{figure}[!htbp]
    \centering
    \includegraphics[width=0.90\textwidth]{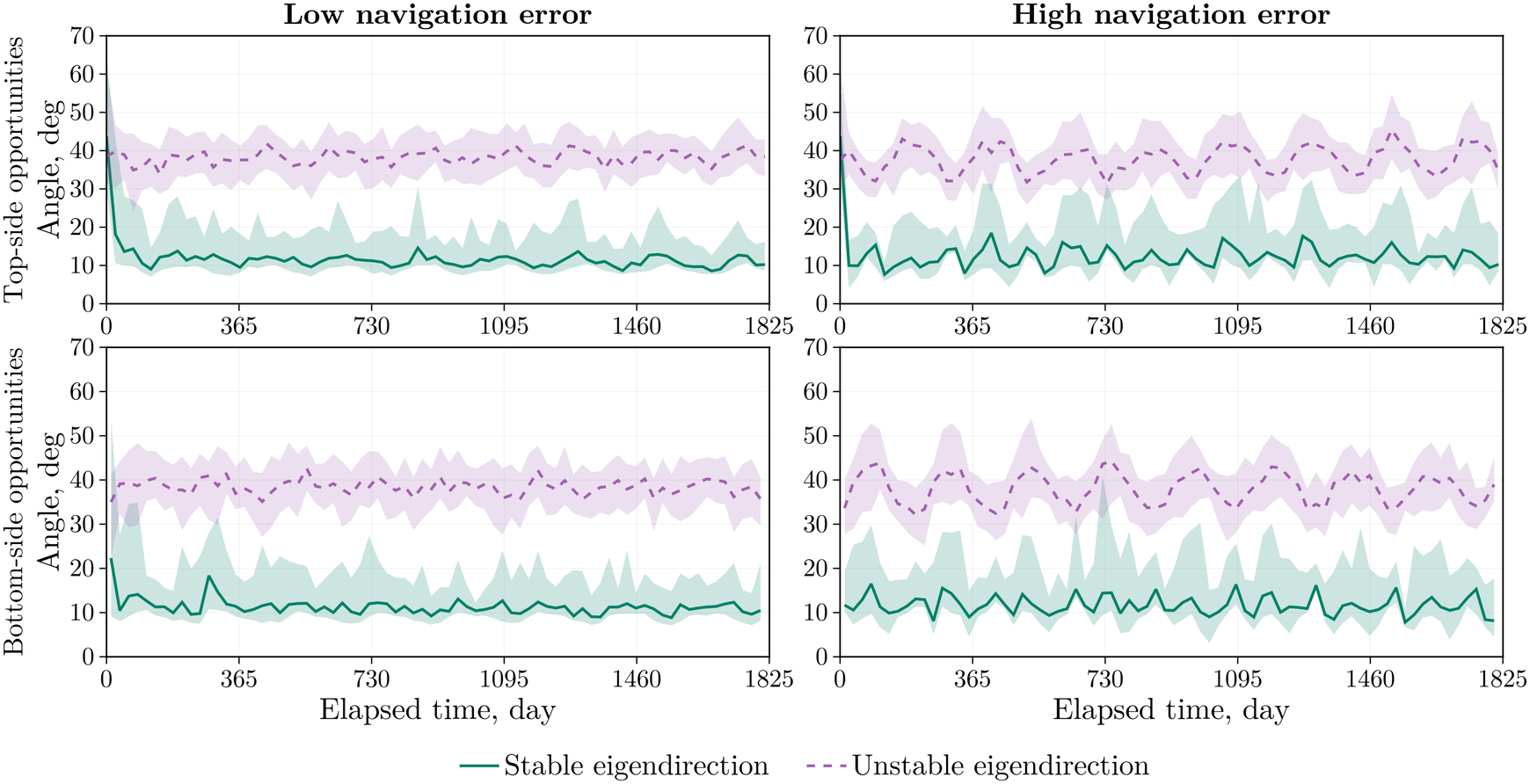}
    \caption{Angles between planned first MPC maneuvers and the corresponding CR3BP stable and unstable velocity axes for low and high navigation error}
    \label{fig:stability_alignment}
\end{figure}

A persistent statistical preference for the stable axis appears in both
navigation error cases and at both the top- and bottom-side control
opportunities. For the high navigation error case, 14,897 of the 14,900 planned first MPC maneuvers exceed the $0.15$~cm/s threshold. The median angle is $12.12^\circ$ to the stable axis and $37.58^\circ$ to
the unstable axis. The top-side medians are $12.40^\circ$ and $37.49^\circ$, and the
bottom-side medians are $11.82^\circ$ and $37.69^\circ$.
Figure~\ref{fig:stability_alignment} shows the maneuver-direction angle
histories; at each opportunity, the curves represent medians across 100
realizations and the shaded regions correspond to the 25th--75th percentile
intervals of the planned first MPC maneuvers that exceed the maneuver threshold. Low navigation error results show the same pattern. For the low navigation error case, 14,871 of the 14,900 planned first MPC maneuvers exceed the $0.15$~cm/s threshold. The median angles are
$11.22^\circ$ to the stable axis and $38.39^\circ$ to the unstable axis. The top-side values are $11.31^\circ$ and $38.32^\circ$, while the
bottom-side values are $11.13^\circ$ and $38.44^\circ$.

The MPC formulation does not include a term that favors either stability
direction. Even so, the median and interquartile histories show a persistent
preference for the stable direction across both navigation-error levels and
at both the top- and bottom-side control opportunities. Individual planned
maneuvers do not always follow this trend.

\subsection{Family-Wide Baseline Generation}

We next apply the same baseline-generation pipeline to 100 members each from the $L_1$ and $L_2$ vertical orbit catalogs. All 200 products are generated using the same mapping from the CR3BP initial
condition to the selected physical epoch, geometry constraints,
multiple-shooting formulation, optimization procedure, and SPK export
workflow, with same-date refinement used where applicable. Each baseline is generated for at least six months and contains a whole
number of revolutions, resulting in durations from $183.6$ to $208.6$~days,
with a median of $194.9$~days.

\begin{figure}[!htb]
    \centering
\begin{subfigure}[t]{\textwidth}
    \centering
    \makebox[\linewidth][c]{%
        \hspace*{0.08\linewidth}%
        \includegraphics[width=0.90\linewidth]{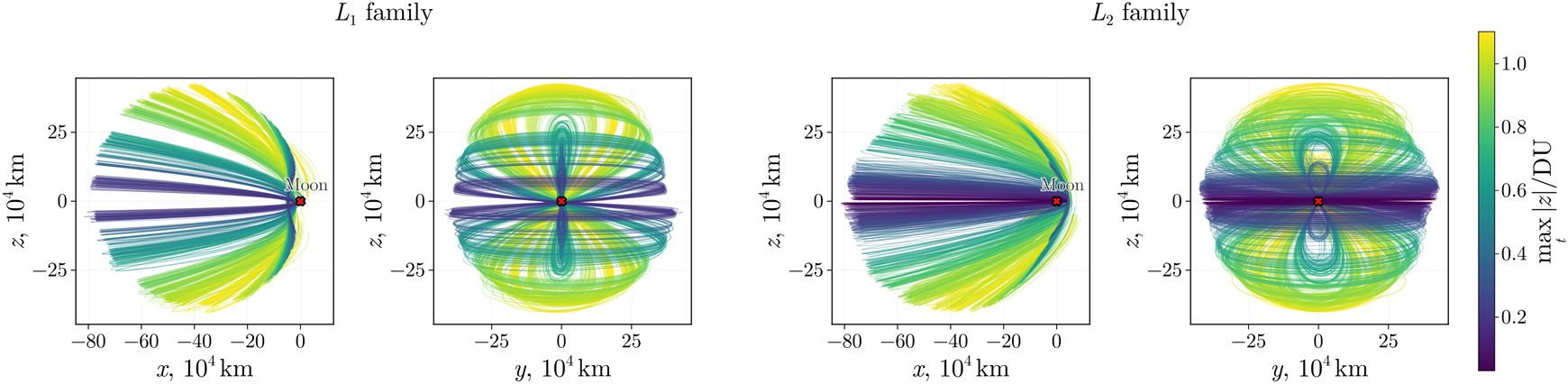}%
    }
    \caption{Family geometry colored by normalized vertical amplitude,
$\mathrm{DU}=389{,}703$~km}
    \label{fig:family_campaign_geometry}
\end{subfigure}

    \vspace{0.25em}
    \begin{subfigure}[t]{\textwidth}
        \centering
        \includegraphics[width=0.90\linewidth]{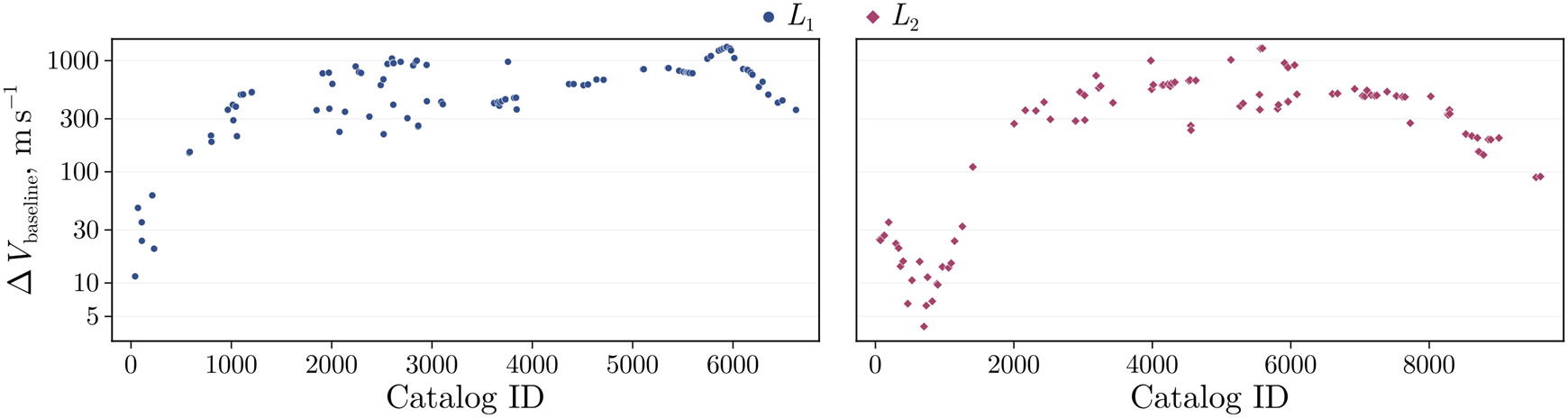}
        \caption{Baseline correction cost}
        \label{fig:family_campaign_cost}
    \end{subfigure}
    \caption{Family-wide six-month baseline results for 100 $L_1$ and 100 $L_2$ catalog members}
    \label{fig:family_campaign}
\end{figure}

For $L_1$, the final correction costs range from $11.5$ to $1327.4$~m/s,
with a median of $606.5$~m/s. For $L_2$, they range from $4.1$ to
$1292.8$~m/s, with a median of $394.4$~m/s. Thus, all 200 short-duration
products require finite deterministic correction at the levels shown, even
though the common generation workflow transfers across the catalog.
Figure~\ref{fig:family_campaign} shows the family geometry and correction-cost
distribution. In Fig.~\ref{fig:family_campaign_geometry}, color denotes the
normalized vertical amplitude $\max_t |z(t)|/\mathrm{DU}$, where
$\mathrm{DU}=389{,}703$~km is the CR3BP distance unit defined above.

\subsection{Selected Long-Duration
\texorpdfstring{$\bm{L_1}$ and $\bm{L_2}$}{L1 and L2}
Vertical Orbits}

\begin{figure}[!htb]
    \centering
    \includegraphics[width=0.80\textwidth]{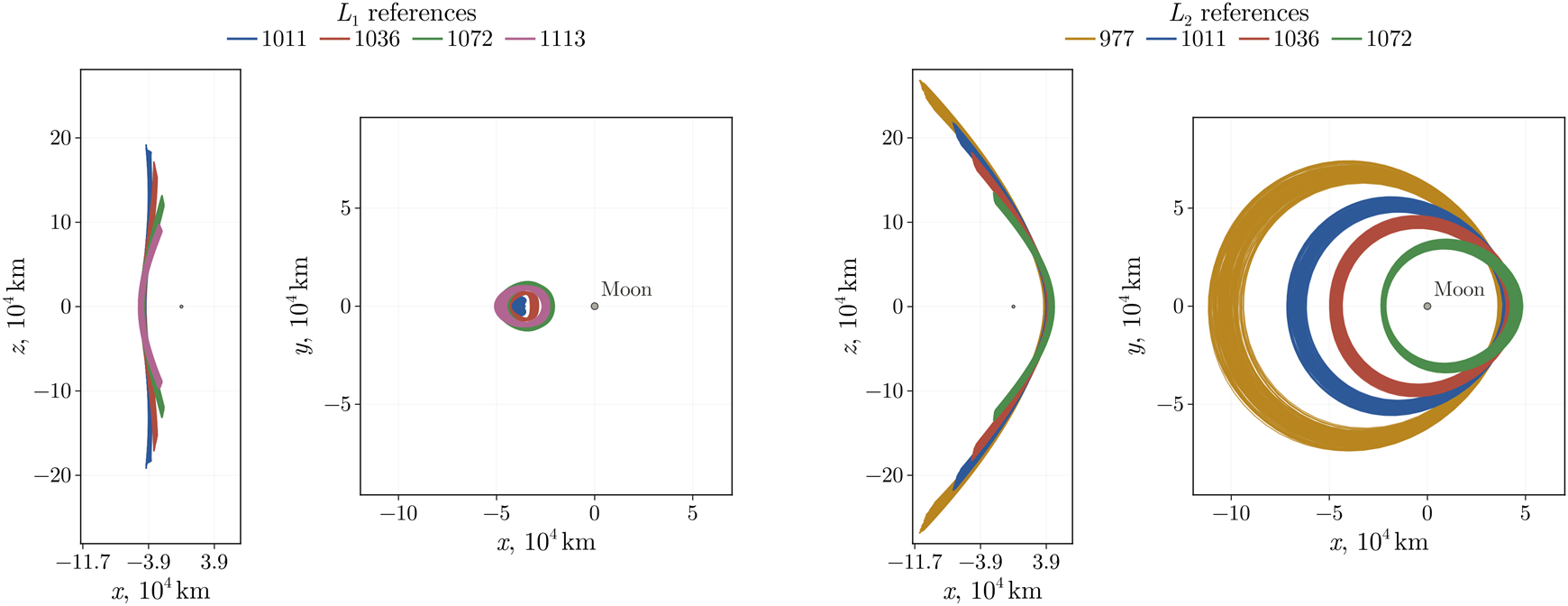}
    \caption{Long-duration high-fidelity SPK reference geometries used in the
station-keeping comparison of selected $L_1$ and $L_2$ vertical orbits}
    \label{fig:eight_reference_geometry}
\end{figure}

For the long-duration comparison, we apply the same receding-horizon MPC
formulation to four $L_1$ and four $L_2$ references, with different vertical amplitudes and in-plane geometries. All eight simulations use $N_{\mathrm{target}}=3$ and
$(\epsilon_r,\epsilon_v)=(25~\mathrm{km},125~\mathrm{m/s})$. These controller parameters could be re-tuned for an
individual orbit of interest, but we retain the same values here to facilitate
a consistent comparison across the tested vertical orbits.
Figure~\ref{fig:eight_reference_geometry} shows the actual SPK references used
in these simulations; the first and last complete revolutions are omitted from
each reference for visualization, and catalog IDs retain the same colors within
each family.

We keep deterministic reference-generation effort separate from closed-loop
maintenance effort. Six references have baseline corrections of order
$10^{-10} \sim 10^{-9}$~m/s. The $L_1$ 1072 and $L_1$ 1011 references require total baseline corrections
of $7.888$~m/s and $328.451$~m/s, respectively, over their complete
approximately 5.5-year durations. For a reference that already contains deterministic correction, the corresponding
mission-level control budget is
\begin{equation}
    \Delta V_{\mathrm{mission}}
    =\Delta V_{\mathrm{baseline,window}}+\Delta V_{\mathrm{SK}}.
    \label{eq:mission_cost}
\end{equation}
Here, $\Delta V_{\mathrm{baseline,window}}$ is the deterministic baseline
correction accumulated during the five-year station-keeping simulation
interval, and $\Delta V_{\mathrm{SK}}$ is the executed station-keeping cost
over the same interval. For $L_1$ 1072, $7.57$~m/s of the complete
$7.89$~m/s baseline correction occurs within this five-year interval.

Among the selected long-duration vertical orbits, seven use two control
opportunities per revolution at the large-$|z|$ extrema. For $L_1$ 1011, we instead use four
control opportunities per revolution while retaining
$N_{\mathrm{target}}=3$ and
$(\epsilon_r,\epsilon_v)=(25~\mathrm{km},125~\mathrm{m/s})$.
The denser cadence is used because this orbit has comparatively small in-plane
motion, which makes it more difficult to control, while its maximum $|z|$
amplitude varies with epoch in the high-fidelity model and contributes to the
larger tracking deviations.

\begin{figure}[]
    \centering

    \begin{subfigure}[t]{0.32\textwidth}
        \centering
        \includegraphics[width=0.98\linewidth]{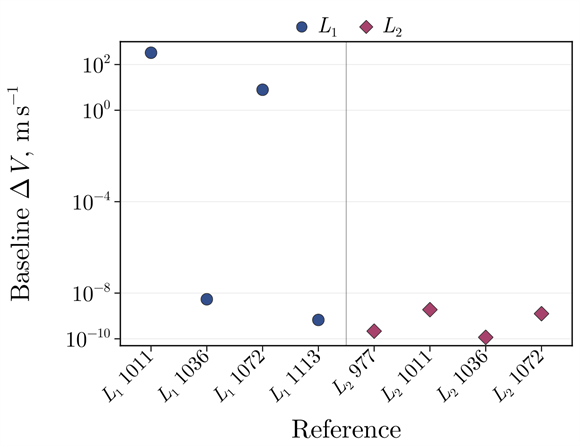}

        \hspace*{0.28\linewidth}%
        \begin{minipage}{0.72\linewidth}
            \centering
            \caption{Baseline correction}
            \label{fig:eight_orbit_baseline_cost}
        \end{minipage}
    \end{subfigure}\hfill
    \begin{subfigure}[t]{0.32\textwidth}
        \centering
        \includegraphics[width=0.98\linewidth]{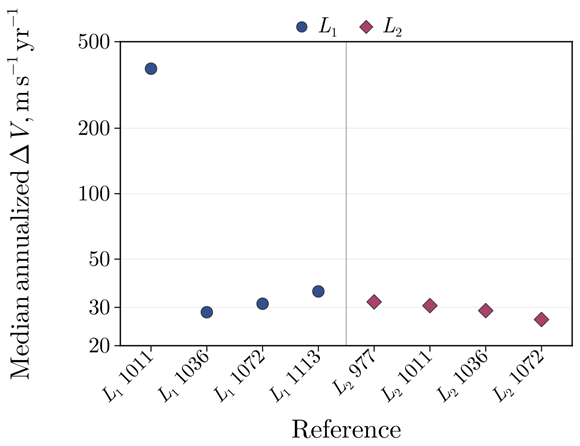}

        \hspace*{0.28\linewidth}%
        \begin{minipage}{0.72\linewidth}
            \centering
            \caption{Station-keeping cost}
            \label{fig:eight_orbit_sk_cost}
        \end{minipage}
    \end{subfigure}\hfill
    \begin{subfigure}[t]{0.32\textwidth}
        \centering
        \includegraphics[width=0.98\linewidth]{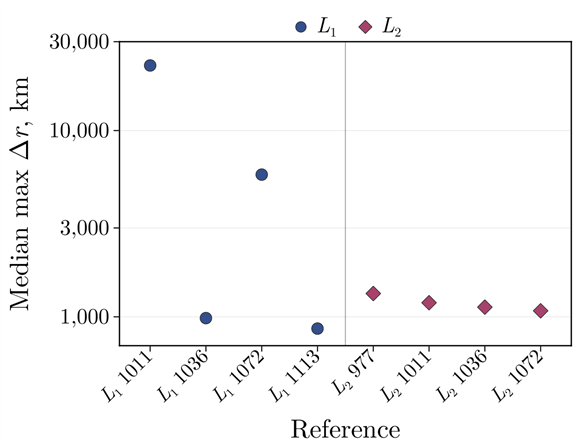}

        \hspace*{0.28\linewidth}%
        \begin{minipage}{0.72\linewidth}
            \centering
            \caption{Maximum $\Delta r$}
            \label{fig:eight_orbit_max_error}
        \end{minipage}
    \end{subfigure}

    \caption{Comparison of complete baseline correction with station-keeping
metrics evaluated under high navigation error for selected long-duration
$L_1$ and $L_2$ vertical-orbit references}
    \label{fig:eight_orbit_synthesis}
\end{figure}

We use the high navigation error case for the selected long-duration comparison,
and all eight references complete 100/100 realizations. The six references with
negligible baseline correction have median annualized station-keeping costs
between $26.386$ and $35.527$~m/s/yr and median maximum position deviations between
$864.6$ and $1333.9$~km. $L_1$ 1072 has a comparable annual station-keeping cost of
$31.185$~m/s/yr but a larger median maximum position deviation of $5801.0$~km. $L_1$
1011 is the clear outlier, with $375.657$~m/s/yr median station-keeping cost and a
median maximum position deviation of $22{,}359.8$~km. Table~\ref{tab:eight_orbit_sk}
lists the exact values and Fig.~\ref{fig:eight_orbit_synthesis} shows the three
metrics side by side.

\begin{table}[]
    \centering
    \caption{Five-year station-keeping comparison under high navigation error for
selected long-duration $L_1$ and $L_2$ vertical-orbit references}
    \label{tab:eight_orbit_sk}
    \begin{threeparttable}
    \centering
    \setlength{\tabcolsep}{4pt}
    \begin{tabular}{@{}ccrrrrr@{}}
        \toprule
        Point & ID &
        \shortstack{CR3BP\\period, day} &
        \shortstack{Stability\\index} &
        \shortstack{Complete baseline\\$\Delta V$, m/s} &
        \shortstack{Median SK,\\m/s/yr} &
        \shortstack{Median max.\\$\Delta r$, km} \\
        \midrule
        $L_1$ & $1011$ & $24.664$ & $176.463$ & $328.451$ & $375.657$ & $22359.8$ \\
        $L_1$ & $1036$ & $23.517$ & $169.600$ & $5.36\times10^{-9}$ & $28.505$ & $984.4$ \\
        $L_1$ & $1072$ & $21.109$ & $177.600$ & $7.888$ & $31.185$ & $5801.0$ \\
        $L_1$ & $1113$ & $17.814$ & $263.734$ & $6.69\times10^{-10}$ & $35.527$ & $864.6$ \\
        $L_2$ & $977$  & $26.221$ & $232.332$ & $2.16\times10^{-10}$ & $31.796$ & $1333.9$ \\
        $L_2$ & $1011$ & $25.436$ & $211.665$ & $1.88\times10^{-9}$ & $30.565$ & $1190.6$ \\
        $L_2$ & $1036$ & $24.501$ & $188.426$ & $1.16\times10^{-10}$ & $28.987$ & $1127.2$ \\
        $L_2$ & $1072$ & $22.195$ & $154.993$ & $1.25\times10^{-9}$ & $26.386$ & $1077.1$ \\
        \bottomrule
    \end{tabular}
    \begin{tablenotes}[flushleft]
        \footnotesize
        \item CR3BP periods and stability indices are from the JPL Three-Body
        Periodic Orbit Catalog~\cite{JPLPeriodicOrbits}.
        \item $L_1$ 1011 uses four control opportunities per revolution; the other
selected long-duration cases use two. The complete $L_1$ 1072 baseline
correction is $7.888$~m/s, of which $7.572$~m/s occurs during the five-year
station-keeping simulation interval.
    \end{tablenotes}
    \end{threeparttable}
\end{table}

The comparison shows why reference generation and station-keeping should not be
collapsed into one metric. $L_1$ 1072 is maintainable at a station-keeping cost
similar to the low-baseline-correction references, but its deterministic
reference already requires several m/s and its tracking envelope is larger.
$L_1$ 1011 is much more demanding in both the reference and closed-loop
problems. Conversely, the six references with negligible deterministic
correction have visibly different geometries yet remain in a comparatively
compact station-keeping cost range under high navigation error.

\section{Conclusion}

We developed and tested an end-to-end workflow that converts an Earth--Moon
vertical orbit from a CR3BP catalog into long-duration high-fidelity references constrained
to retain the broad CR3BP geometry and then maintains those references with receding-horizon MPC. For $L_2$ 1036, the initial solve produces a 5.5-year reference requiring only
$1.16\times10^{-10}$~m/s of deterministic baseline correction, and the
segmented SPK reproduces the optimized coast arcs to the reported numerical
precision. By uniformly varying the half-widths of the rotating-frame geometry boxes, we
evaluate how the allowed deviation from the CR3BP geometry affects the
deterministic baseline correction. At the tightest tested box scale, the
geometry constraints are active throughout the reference and require
$12.50$~km/s of correction, whereas sufficiently wide boxes become inactive
and recover an effectively ballistic reference.

The start-epoch sweep provides a separate robustness result. Across 90
consecutive start dates from January through March 2026, we find local optima on all 90 cases; 37 require no refinement, 44 require one
warm start, and 9 require two or more. Because the amount of refinement differs
between dates, the final correction curve is descriptive and is not used to
claim a controlled physical epoch-cost relationship.

A properly tuned MPC configuration maintains $L_2$ 1036 in all 200 five-year Monte Carlo runs under low and high navigation error.
Within the tested uncertainty levels, navigation error has the largest effect
on maneuver cost and tracking dispersion. The planned first MPC maneuvers also show
a persistent ensemble-level preference for the corresponding stable CR3BP
velocity axis over the unstable axis, even though no such alignment is imposed
in the OCP. Together with the finite-time Lyapunov exponent along the actual
high-fidelity reference, this provides a complementary dynamical interpretation of the closed-loop behavior.

The same baseline-generation workflow generates references for all 200
approximately six-month family cases. The selected long-duration
station-keeping comparison shows that deterministic reference-generation
effort and closed-loop maintenance effort can differ substantially. Among the selected long-duration $L_1$ and $L_2$ vertical-orbit references,
six require only negligible deterministic correction and cluster in a relatively compact station-keeping cost range under high navigation error; $L_1$ 1072
and especially $L_1$ 1011 show that a maintainable reference can still be
operationally expensive or difficult to track. Reference design and station-keeping performance should therefore be
evaluated separately and then combined at the mission level.

\balance
\bibliographystyle{AAS_publication}
\bibliography{references}

\end{document}